\documentclass[useAMS,usenatbib]{mn2e}
\usepackage{graphicx}
\usepackage{longtable}
\title[Triggered star formation beyond the solar circle]{SOAR-OSIRIS 
observations of the Sh 2-307 HII region: Triggered 
star formation beyond the solar circle}
\author[A. Roman-Lopes]{A. Roman-Lopes$^{1}$\thanks{roman@dfuls.cl}\\
$^{1}$Physics Department - Universidad de La Serena - Cisternas, 1200 - La Serena - Chile }

\begin{document}

\date{}

\pagerange{\pageref{firstpage}--\pageref{lastpage}} \pubyear{2002}

\maketitle

\label{firstpage}

\begin{abstract}
   This work aims to the study of the Sh 2-307 HII region and related stellar population.
   Near-infrared imaging and spectroscopic observations in the direction of Sh 2-307 were performed using OSIRIS at SOAR Telescope.
   The photometric data were analysed from color-color, and colour-magnitude diagrams, while the spectroscopic
   results were interpreted from the comparison of the science spectra with that obtained from known OB stars.
   From J-, H- and K-band spectra of the brightest source in the cluster, we
   conclude that it has a near-infrared spectra compatible with that taken for O9{\sc v}-O9.5{\sc v} stars. 
   Using the derived spectral type and the respective J, H and K-band photometry, we compute a heliocentric distance 
   of 3.2$\pm$0.5 kpc, which for R$_0$ = 8 kpc, puts this cluster at more than 10 kpc from the Galactic centre.
   From the Br$\gamma$, H$_2$, and continuum narrow-band images we
   were able to detect both the NIR counterpart of the associated HII region, as well as,
   the interface between the ionised and the cool molecular gas.
   Using the 5 Ghz continuum flux density taken from the PMN catalogue and the Br$\gamma$ narrow band image we estimate 
   that the HII region has a mean diameter of 0.94$\pm$0.15 pc, mean electron density of 550 cm$^{-3}$ and an
   estimated dynamical age of 1.6$\times 10^6$ years.
   The large fraction of sources presenting excess emission at 2$\mu$m suggests that the stellar population is 
   very young, with many sources still in the pre-main sequence accreting phase. 
   By the use of theoretical pre-main sequence tracks we derived a cluster mean age of about 2.5 Myears, and from the
analyses of the fraction of excess emission sources as a function of their spatial distribution we found 
evidence for an age
   spread for the embedded pre-main sequence stellar population.
   Finally, from the study of the spatial distribution of the low-mass sources relative to the main-cluster source and
   associated photo-dissociation zones, we conclude
   that the O-type star probably has been triggering the star formation process in the region.

\end{abstract}


\section{Introduction}
The studies of Galactic stellar clusters are important to improve our knowledge
of the physical processes leading to their formation and evolution.
The stars in a cluster share the same parental molecular cloud, are formed
more or less simultaneously in time, and contain samples of stars that span a wide range of 
stellar masses within a relatively small volume of space. 
In this sense, they can be used to provide tests for the stellar evolution theory.
As an example, we notice that such systems are the smallest physical 
entities in which one can perform a meaningful determination of the 
stellar initial mass function (IMF) (Salpeter, 1955; Scalo, 1986).
On the other hand, because young stellar clusters are excellent tracers of the spiral 
pattern of external galaxies, the study of the spatial distribution of the Galactic clusters
play a key hole in the understand of the structure and evolution of our own galaxy.

    \begin{figure}
   \includegraphics[bb=14 14 1322 3898,width=6 cm,clip]{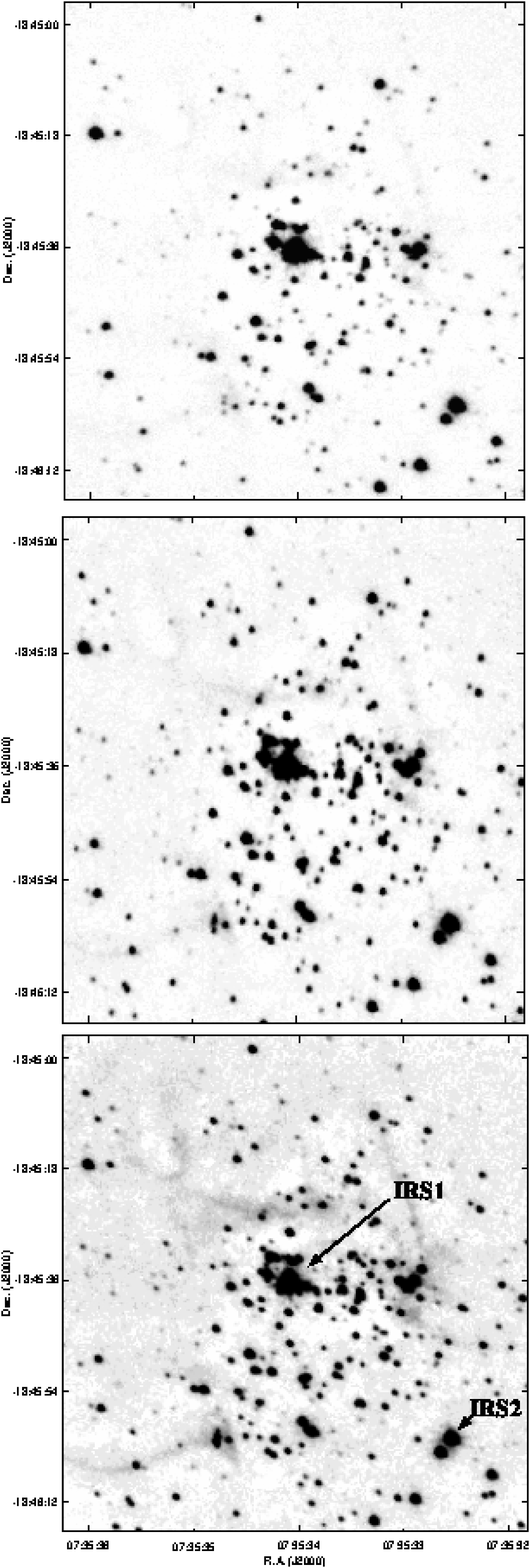}
      \caption{The near-infrared $J$ (top), $H$ (centre) and $K_S$ (bottom) OSIRIS (70\arcsec $\times $ 60\arcsec) 
	  images of the region in the direction of the Sh 2-307 HII region. North is to the top, east to the left.
              }
         \label{FigVibStab}
   \end{figure} 
 
Sharpless 307 (Sh 2-307), is a small HII region located in the outer 
Galaxy at l = 234.58, b= +0.84 (Sharpless, 1959). 
It was detected by the Parkes-MIT-NRAO (PMN) 4.85GHz Survey as a radio source measuring
about 1.3$\arcmin \times 1.2\arcmin$, with integrated flux density $S_{5Ghz}$ of 726$\pm$39 mJy.
It is part of a complex that was studied by Russeil et al. 
(1995), who from Fabry-Perot H$\alpha$ interferometric observations,
concluded that the HII regions can be separate in two velocity groups.
The first group with $V_{LSR}\sim$ 16 km s$^{-1}$, is located at a kinematic distance of 2.2 kpc
while the other 
with $V_{LSR}\sim$ 43 km s$^{-1}$, corresponds to a kinematic distance of 4.2 kpc.
From this scheme, the Sh 2-307 HII region (with a radial velocity of 37 km s$^{-1}$) would belongs to the 
second molecular cloud. However, its radial velocity is somewhat lower than that for the other HII regions, 
and as suggested by the authors, a possible explanation for the discrepancy 
between the velocities of the ionised and molecular material, would come from the effects 
of a "Champagne flow" (Tenorio-Tagle, 1979), which would be produced by the ionization front of the associated
HII region located at the near edge of the molecular cloud. 
The presence of a far-IR source (IRAS07333-1838) that has colours characteristic of ultra-compact 
HII regions (Wood \& Churchwell, 1989), and the detection of a 22 GHz H$_2$O maser by Han et al. (1998), 
suggest that star formation is still occurring there.

Moffat et al. (1979) were the first group to obtain some photometric measurements of point sources
in the region. They performed $UBV$ photometry of five objects in the 
direction of the complex that contains the Sh 2-307 HII region. However, the only point source directly 
related with this HII region
is their source \#3 (which has $V$ = 12.47, $B-V$ = 0.79, $U-B$ = -0.32). 
The other point sources are placed outside the core of the Sh 2-307 HII region, 
not belonging to the same star forming complex. Hunter \& Massey (1990) performed optical spectra for a bright optical
source in the core of the Sh 2-307 region, concluding that it might be a B0{\sc v} star.
Dutra et al. (2003), from the analyses of the superficial density of sources in the 2MASS 
image survey, suggested the presence of a cluster
of stars in the direction of Sh 2-307 (identified in their catalogue as cluster candidate \#7). 
Unfortunately, the poor spatial resolution of the 2MASS images do not able one to properly resolve the stars in the inner part of the system.

In this paper we report the results of a near-infrared spectrophotometric survey made in the direction
of the core of the Sh 2-307 HII region. For the first time we were able to fully access its stellar
content, identifying the main-ionizing source of the cluster and addressing its stellar 
population.  
In section 2 we describe the observations and the data reduction procedures, in section
3 we present the results, in section 4 the analyses and discussion, and in section
5 we present our conclusions.

\begin{figure*}

   \includegraphics[width=9cm]{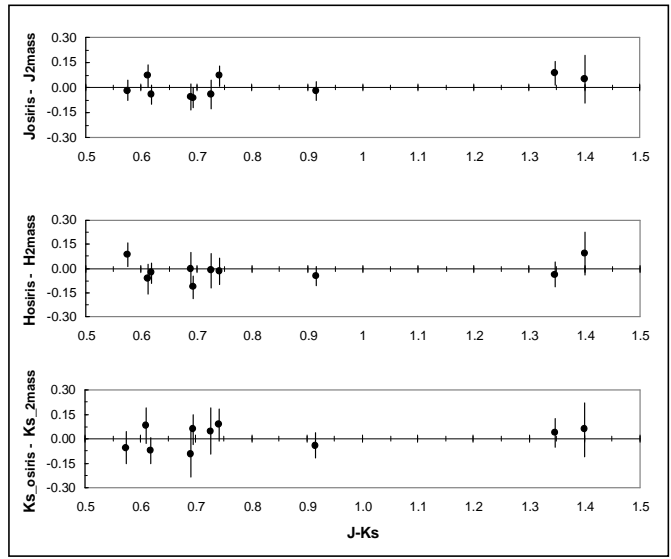}
      \caption{The difference between the OSIRIS and 2MASS J-, H- and K$_s$-band photometry as function of the $J-K_s$ colours. 
	  We can see that there 
	  is no need to apply colour correction terms.}
         \label{FigVibStab}
   \end{figure*}

\section{Observations and data reduction}
In this work we used the Ohio State Infrared Imager/Spectrometer (OSIRIS\footnote{OSIRIS is a collaborative project between the Ohio State University 
and Cerro Tololo Inter-American Observatory (CTIO) and was developed through NSF grants AST 90-16112 and AST 92-18449. CTIO is part of the National 
Optical Astronomy Observatory (NOAO), based in La Serena, Chile. NOAO is operated by the Association of Universities for Research in Astronomy (AURA), 
Inc. under cooperative agreement with the National Science Foundation.}), coupled to the 4.1 m telescope of the Southern Observatory for Astrophysical 
Research (SOAR). The observatory is placed at Cerro Pachon, Chilean Andes (about 2800 m above the sea level). It was constructed by the consortium of 
the Brazilian Ministry of Science, the National Optical Astronomy Observatories, the University of North Carolina and Michigan State University.

\subsection{OSIRIS imaging observations}

High resolution near-infrared (NIR) imaging observations were performed in the direction of Sh 2-307 ($\alpha$ = 7h35m34.19s and  
$\delta$ = -18d45m35.6s, Equ J2000). The data were taken during the 2007 November, 8 night, using the OSIRIS's f/7 camera, 
which at the SOAR Telescope 
produces images with a plate scale of about 0.14$\arcsec$ per pixel. We used the $J,H$ and $K_S$ broad-band filters and the $H_2$, Continuum 
(at 2.14$\mu$m) and $Br\gamma$ narrow-band filters.

From the OSIRIS acquisition images of the region, we detected the presence of extended emission.
In order to perform a good subtraction of the thermal background component from the science frames, it is necessary to construct a "clean" median 
combined image, free of any residual of point sources, as well as, of 
extended sources. The FOV of the OSIRIS camera in the f/7 mode is about 80$\arcsec$, which is only a bit larger than the apparent 
angular size of the cluster, which was estimated as 1$\arcmin$ by Dutra et al. (2003). 
This feature could limit our capability to construct useful "sky" frames when using only the science ones. 
We avoid this problem by performing
"on source", and "sky" sequential observations (source-sky-source-sky...). The "sky" frames were taken in a region 
located about 2$\arcmin$ east from the cluster centre, with coordinates $\alpha$ = 7h35m42.46s and $\delta$ = -18d45m41.7s (J2000) .

Despite the extra amount of time necessary to perform "on-source" and "sky" alternate 
observations, there are some advantages when using this procedure.
From the "sky" frames, we constructed $J, H$ and $K_s$ images
that enabled us to perform photometric calibration of the OSIRIS data, using the results 
from the 2MASS point source catalogue. 
On the other hand, under the assumption that the point sources in the "sky" region are 
representative of the "local" Galactic field stellar 
population, the analyses of their magnitudes and colours, provide additional clues that can 
help us to identify and exclude the sources that probably do not belong to the star forming 
region. 

We obtained a set of five images per filter for both, the sky and cluster coordinates. 
This was done following a cross pattern in which the telescope 
was nodded by 8$\arcsec$ in the NSEW directions (always from the associated central coordinates). 
The individual exposure times 
for the broad- and narrow-band frames were 60s, and in order to guaranty good photometry 
also for the brightest sources (avoiding saturation), 
we also took a set of J-, H- and K$_S$-band short exposure time (5s) images. In order 
to correct the raw frames by the differences in the pixel
response, a set of on-off flat-field images were taken for each broad- and narrow-band filter
we used.

The raw frames were reduced using the CTIO infrared reduction 
(CIRRED\footnote{http://www.ctio.noao.edu/instruments/ir\_instruments/datared.html}) 
package within IRAF\footnote{IRAF is distributed by the National Optical Astronomy 
Observatories, which are operated by the Association of Universities 
for Research in Astronomy, INC, under cooperative agreement with the National Science 
Foundation}, following the NIR reduction procedures described in Roman-Lopes et al.
(2003).
The final broad- and narrow-band NIR images for the cluster and "sky" (hereafter named 
"control") regions, were constructed by shifting and combining 
the reduced raw frames. 
The mean values of the full width at half maximum (FWHM) of the point sources in the 
J-, H- and K$_S$- band combined images are 
0.75$\arcsec$, 0.70$\arcsec$ and 0.64$\arcsec$, respectively.

 \begin{figure*}
 \centering
  \vspace{20pt}
   \includegraphics[bb=36 433 462 776,width=12.5cm,clip]{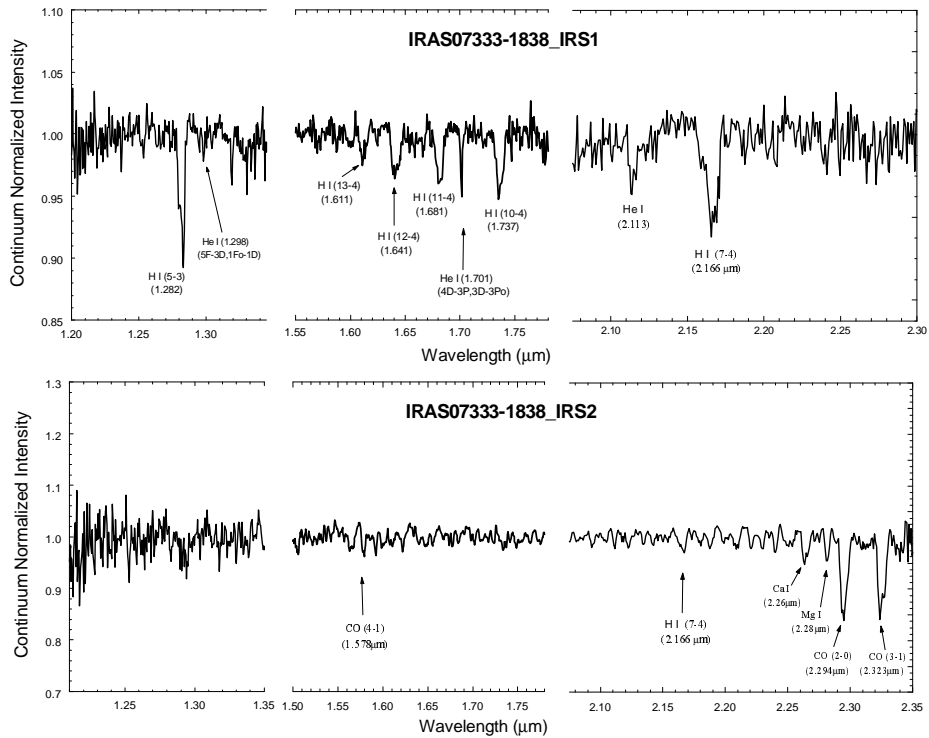}
      \caption{The NIR spectra for the IRS1 and IRS2 sources. 
	  The spectral resolution in the $K$ band is R$\sim$1200 and the 1$\sigma$ uncertainty in the wavelength 
	  calibration is about 7 \AA. 
	  The mean signal to noise ratio is 120 for IRS1 spectra and about 50 for IRS2 one.
	  The spectra of IRS1 source shows the Hydrogen and Helium recombination lines typical of hot stars, while that for IRS2, 
	  presents CO and metallic lines commonly 
	  found in late-type stars and young stellar objects.}
         \label{FigVibStab}
   \end{figure*}

\subsection{NIR spectroscopic observations}

The NIR spectroscopic observations of the two brightest 
sources in the cluster region were made during the 
2007 November, 24 night. The J-, H- and K-band spectra were taken using OSIRIS in the Low-Resolution 
(R $\sim$ 1200) multi-order cross-dispersed (XD) mode. In the XD mode, 
the instrument operates using the f/2.8 camera and the short slit 
(27\arcsec $\times$ 1\arcsec), covering the three bands simultaneously 
in adjacent orders. 
The spectra were acquired using the standard AB nodding technique (with an offset of 12\arcsec). 
Individual exposure times at each nod position was 120 s. For each source, we obtained 
16 individual frames that resulted in a total individual 
exposure time of 32 minutes. In order to remove telluric atmospheric 
absorption effects from the science spectra, nearby A and G spectroscopic standard stars 
were observed at similar airmasses, before and after the 
science targets.

    \begin{figure}
   \centering
   \includegraphics[bb=14 14 302 533,width=6cm,clip]{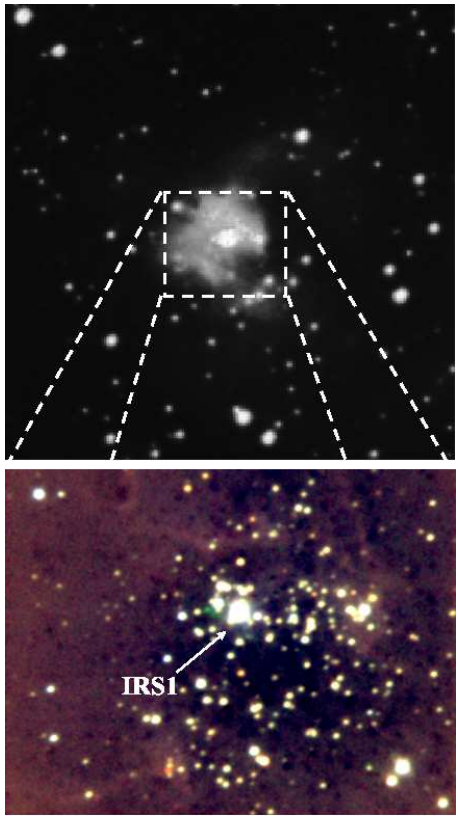}
     \caption{The DSS R-band image (about 5 arcmin in size)  of the core of the SH 2-307 HII region (top), 
	 and the J- (blue), H- (green) and K$_S$-band (red) three colour composite image 
	 ($\sim$ 70\arcsec$\times$ 50\arcsec)
	 of the section delineated by the white dotted square in the R-band image.
	 The IRS1 source for which a NIR spectra were taken, is indicated. Note
	 the large quantity of sources around the main source. A cavity inside the cloud (delineated by 
	 the K-band emission) is clearly seem. In this figure, North
	 is to the top and east is to the left.}
         \label{FigVibStab}
   \end{figure}

The spectroscopic data were reduced using the CIRRED package within IRAF.
The two-dimensional frames were sky-subtracted for each pair of
images taken at the two nod positions A and B, followed by division
of the resultant image by a master flat. 
Thereafter, wavelength calibration was applied using the sky lines;
the typical error (1-$\sigma$) for this calibration was $\sim$7~\AA.
The multiple exposures were combined, followed by
one-dimensional extraction of the spectra. Telluric atmospheric
correction using the spectroscopic standard stars completed the
reduction process. 
In this last step, we divided the target spectra by the spectrum of the A0{\sc v} spectroscopic standard star,
already free of photospheric features. In this particular case, the Hydrogen series are the main features
present in the {\it J, H}- and {\it K}-bands. The $J$ and $K$ band hydrogen lines in the 
spectra of the A0V spectroscopic 
standard star, were carefully removed by interpolation across their wings, using the continuum points on either
side of the line.

In the case of the H-band, the subtraction of the hydrogen absorption lines can not be successfully made directly 
from the spectrum of the A0{\sc v} telluric star. The relative position 
of the multiple lines of the Bracket series and the presence of some strong telluric features, do not
permit a good fitting of the adjacent continuum, which is fundamental to remove the line profiles. 
In order to assure good results, we removed the Hydrogen lines from the A0{\sc v} H-band spectrum using the procedure 
applied by Blum et al (1997). 
The method consists in to obtain a good fitting of the hydrogen
lines, using the spectrum obtained by the division of the A standard star spectrum, by that of the G star. In the case 
of the G star spectrum,
its own H{\sc i} intrinsic lines (which in general are weaker than that found in the A0{\sc v} spectra), were previously
removed by hand using as template, the NOAO solar atlas of Livingston \& Wallace (1991).
The last step was to corrected the A0{\sc v} spectrum by that of the model, obtaining a H-band spectrum free of the 
hydrogen Bracket lines.

Finally, in order to assure good cancellation of the telluric bands, the task
{\it telluric} of IRAF was employed. The algorithm interactively
minimises the RMS in specific sample regions, by adjusting the shifts
and scales between the standard and science spectrum to best divide
out telluric features present in the latter. The shifting allows
for possible small errors in the dispersion zero-points.
The intensity scaling allows for differences in the airmass and
variations in the abundance of the telluric species. Typical
values of the shifts were equivalent to a few tenth of a pixel (about $\sim$2\,\AA), 
while the scaling factors were not larger than 10\%.

\subsection{The NIR spectra of known OB stars}

In order to estimate the spectral type of the brightest source in the cluster, its spectra
were compared with that obtained from the known OB stars (hereafter named $templates$), which are present in Table 1. 
The $J$, $H$ and $K$ band spectroscopic observations of the template stars were performed in different nights,
in which all the spectra were acquired 
from the same equipment used during the observations of the Sh 2-307 sources 
(e.g. same telescope, instrument setup, and technique described in Section 2.2).
It means that all spectra have virtually the same spectral resolution 
(that can change a bit due to differences in the seeing conditions, airmass, focus, etc), and cover the 
same spectral range. Due to the brightness of the template stars (all sources have 
$K$ magnitudes less than 8), the signal-to-noise ratio of the resulting spectra are 
150 in excess.

\section{Results}

\subsection{NIR photometry}     

The J-, H- and 
K$_S$-band images of the core of the Sh 2-307 HII region (hereafter named $cluster$ region) are shown in Figure 1. 
The NIR photometry of the point sources in the cluster and control regions 
were performed using the DAOPHOT (Stetson, 1987) package 
within IRAF. The selection of the point sources in the NIR images, was made using the task 
daofind, assuming a detection threshold of 10$\sigma$ 
above the local sky. Due to source confusion in some crowded parts of the cluster region, 
we choose to perform point spread function (PSF)
photometry. The procedure we applied was that described 
in Roman-Lopes \& Abraham (2004), using a PSF radii
of 14 pixels for the J- and H- bands and 12 pixels for the K$_S$-band.

The zero-points for the J-, H- and K$_S$-band OSIRIS photometry were determined using the 
results from the 
2MASS\footnote{http://irsa.ipac.caltech.edu/applications/2MASS} all sky point source 
catalogue (PSC). For each broad band filter, a set of isolate 
stars in the control images was used to calibrate the OSIRIS 
photometric data. Only stars with good 2MASS photometry 
(flags A and B) were used. 
The mean values of the errors in the zero point magnitudes are 0.03, 0.05 and 0.06 mag 
for the J-, H- and K$_s$-bands, respectively.
The final photometric errors were calculated as the quadratic sum of the mean zero-point 
errors mentioned above and the associated photometric 
uncertainty for each source, as given by the ALLSTAR routine.
The values of the completeness limits for J-, H-, and K$_S$-bands are 17.6, 17.2, and 16.7 
mag, respectively, and were derived from the point at
which the number of detected sources of magnitude m, N(m) deviates from a straight line 
in the log (N) versus m diagram (Roman-Lopes et al. 2003).

We found a good correlation between the OSIRIS photometry and that from the 2MASS 
survey. 
The plots with the differences between the OSIRIS and 2MASS J-, H- and K$_S$-band 
magnitudes as function of the $J-K_s$ colours are shown in Figure 2. 
We can see that there is no need to apply colour-term correction in the derived OSIRIS 
magnitudes.
 The J-, H- and K$_S$-band photometry obtained for all sources detected in the 
 cluster region are shown in Table 2 (which is available in its entire form in the
 electronic version of the article).   
 
\begin{figure*}
\centering
  \vspace{20pt}
   \includegraphics[bb=42 86 373 777,width=9.86cm,clip]{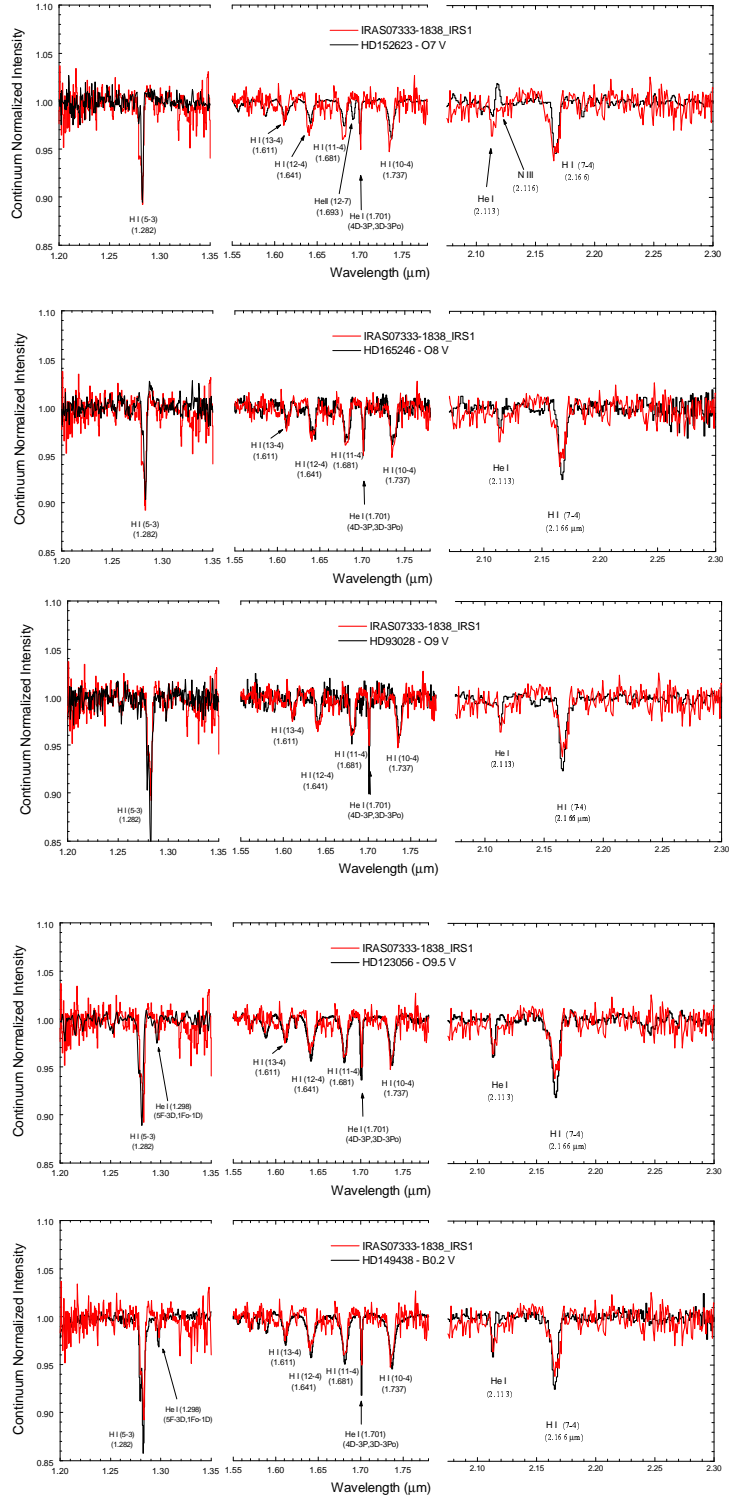}
      \caption{The NIR spectra of the IRS1, superimposed on the J-, H- and K-band spectra of the known OB stars, HD152623 (O7{\sc v}), 
	  HD165246 (O8{\sc v}), HD93028 (O9{\sc v}),
	   HD123056 (O9.5{\sc v}) and HD149438 (B0.2{\sc v}).
	  }
         \label{FigVibStab}
   \end{figure*}
  
 \subsection{NIR spectra of IRS1 and IRS2 sources}

Figure 3 shows the $J, H$ and $K$ band spectra taken for IRS1 and IRS2, the two brightest sources in the cluster region
(see Figure 1). 
The spectra of the IRS1 source present features typical of hot stars. In fact, there are several
Hydrogen recombination lines from the Bracket series at 1.611$\mu$m, 1.641$\mu$m, 1.681$\mu$m, 
1.737$\mu$m and 2.166$\mu$m, and also the H{\sc i} (5-3) Pachen line (at 1.282$\mu$m), the last one in the $J$ band.
We also found the He{\sc i} lines at 1.701$\mu$m and 2.113$\mu$m, which  
indicate that IRS1 must be hotter than a B2{\sc v} star (Hanson et al. 1996).

The NIR spectra of the IRS2 source is completely different from that taken for IRS1. There are no hydrogen or He{\sc i} lines in the $H$ band, and we found only a week Br$\gamma$ absorption line in the $K$ band. On the other hand, we can see the CO lines at 1.578$\mu$m, 2.294$\mu$m and 
2.323$\mu$m, which are commonly seem in late-type stars (Ali et al. 1995; Ivanov et al. 2004), as well as, also found in the NIR spectra of young stellar objects (Casali \& Matthews 1992, Hoffmeister et al. 2006). 
In summary, the K-band spectrum of IRS2 is not characteristic of hot stars, but considering its aparent location in the densest part of the cloud and its spatial correlation with some extended emission seem in the R-band image, we can not dischard its condition as a member of the cluster, being a candidate to intermediate mass young stellar object (YSO).
 
\section{Analyses and discussion}

\subsection{The spectral type of main cluster source}

As already mentioned before, Hunter \& Massey (1990) obtained optical spectroscopy data 
for a non-resolved source in the direction of the core of the Sh 2-307 HII region, classifying it as a B0{\sc v} star. 
Thanks to the high-resolution NIR images obtained by the SOAR telescope, for the first
time we resolved the sources in the central part of the HII region, separating the main ionizing agent of the cluster from its less luminous companion.
In Figure 4 we show the $R$
band image of the region (taken from the DSS archive), and the corresponding (at least
for the central part) OSIRIS combined three colour image. 
One can see a large quantity of point sources in the inner part of the region, with IRS1 
emerging as the brightest one. The $K$ band image (represented in red) clearly shows the existence of a cavity inside the cloud,  
generated by the ionizing wind of the central source of the HII region.

In Figure 5 we present the IRS1 NIR spectra superimposed on that obtained for the
five known OB stars (Maiz-Apellaniz et al., 2004, Hanson et al., 2005)  present in \textbf{Table 1}.
From the comparison of the IRS1 spectra with that of the templates stars, 
based on the presence and strength of the HeI lines, and the lack of HeII lines, we conclude 
that IRS1 probably is an O9{\sc v}-O9.5{\sc v} star.

The spectral type estimate we found for the main source in the region is only a bit different 
(in terms of sub-types) than
that obtained by Hunter \& Massey (1990).
However, we notice that this small spectral type change imply in a 
significant difference in luminosity, with large impact in the quoted distance. 
In fact, the difference between visual absolute magnitudes of B0{\sc v} and O9.5{\sc v} stars, 
may be as high as 1.2 mag (Wegner 2006). 
Yet if we consider the case of zero age main-sequence stars (ZAMS), the 
differences in terms of absolute visual magnitudes drops to about 0.6-0.8 (Hanson et al. 1997), a value still 
considerable.

\subsection{Spectrophotometric distance of the IRS1 source: An estimate for the cluster distance}

\begin{figure*}
\centering
\vspace{18pt}
   \includegraphics[bb=38 443 290 758,width=9cm,clip]{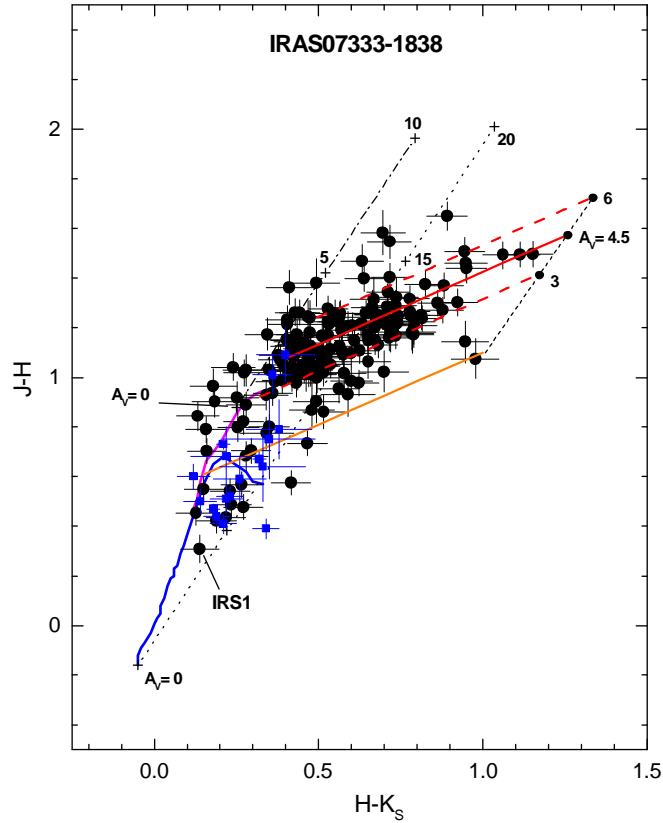}
      \caption{The $J-H$ versus $H-K_S$ diagram of the sources detected in the direction of 
	 IRAS07333-1838 source (represented by the black filled circles) and that from the control region (the blue filled triangles).
	 The theoretical locus for the main-sequence stars is represented by the blue
	 continuum line, while that for the giant branch (Koornneef, 1983) is represented by the pink continuum line.
	 We also represented the non-reddened theoretical locus for the classical T-Tauri, by the orange line. The corresponding 
	 reddening vectors are indicate respectively by the dashed-dotted, dotted, and dashed lines. We also represented 
	 by red lines, the theoretical CTT locus reddened by 3, 4.5 and 6 magnitudes of visual extinction 
	 (Fitzpatrick, 1999).}
         \label{FigVibStab}
   \end{figure*}

From the spectral type estimate obtained in the previous section, and the NIR photometry, 
we are able to compute the distance to the main cluster source.

The NIR absolute magnitudes were computed using the intrinsic colours for early-O stars given by
Koornneef (1983), and the
absolute $M_V$ magnitudes for ZAMS O9{\sc v} and O9.5{\sc v} stars given by Hanson et al. 1997. 
We choose to use the absolute magnitude values for ZAMS in order to be conservative
in the distance determination. In fact, it is well known that ZAMS OB dwarfs 
found in very young clusters are less luminous than that found in the Galactic
field (Hanson et al. 1997, Walborn 2002, Wegner 2006). 
The measured $J$, $H$ and $K$ magnitudes for IRS1 (presented in Table 1) are 
10.66$\pm$0.05, 10.35$\pm$0.06 and 10.21$\pm$0.07 while its $(J-H)$ and $(H-K)$ colours
are 0.31$\pm$0.08 and 0.14$\pm$0.09, respectively. The mean
intrinsic $(J-H)$ and $(H-K)$ colours for O9{\sc v}-O9.5{\sc v} stars are $(J-H)_0$=-0.12 and $(H-K)_0$=-0.05 (Koornneef 1983).
The resulting mean colour excess are $E(J-H)$=0.43 and $E(H-K)$=0.19, which using the standard 
interstellar extinction law given by Fitzpatrick (1999) correspond to $A_V$=3.6$\pm$0.7 magnitudes 
or $A_J$=1.1$\pm$0.2 and $A_K$=0.4$\pm$0.1.

Using the appropriate values, we computed a distance of 3.2$\pm$0.5 kpc, which for R$_0$= 8 kpc, puts the  
Sh 2-307 HII region at more than 10 kpc from the Galactic centre.
The quoted error in the value for the distance to IRS1 source was calculated 
taking into account the associated uncertainty in the extinction, photometry, and also the error 
that comes from the uncertainty on the absolute magnitude derived from
the spectral type determination itself (for example, $M_J$=-3.0 for an O9{\sc v} star and -2.85 for 
an O9.5{\sc v}).
As a last comment, we notice that in the scenario where IRS1 appears as a main-sequence star (e.g. not a ZAMS), 
the quoted distance would increase by about 1.1 kpc.

\subsection{The stellar population of the Sh 2-307 HII region} 

\subsubsection{The colour-colour diagram: A cluster of PMS stars}

\begin{figure*}
\centering
\vspace{30pt}
\includegraphics[width=7cm,clip]{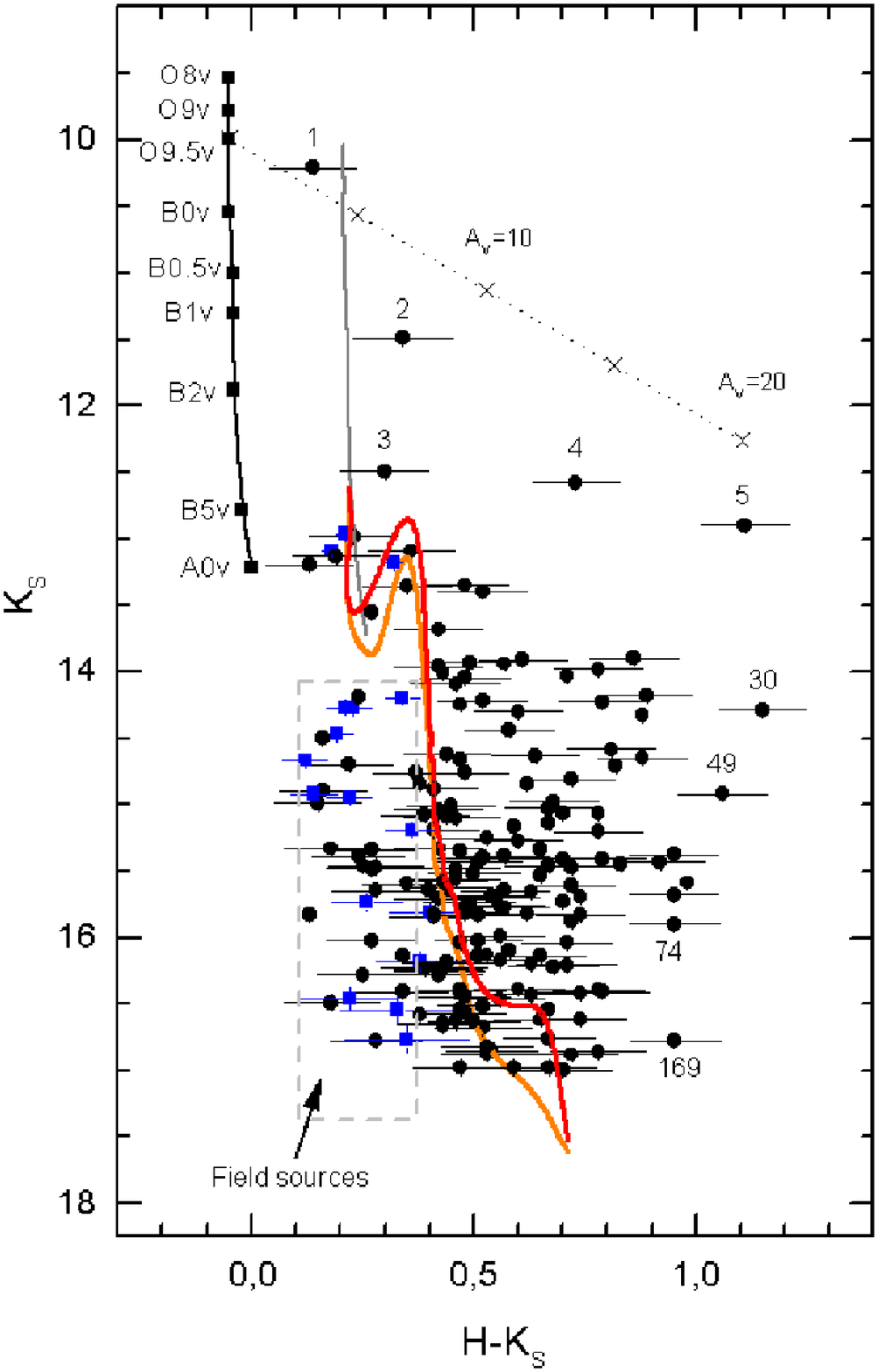}
\caption{The $J$ versus $H-K_S$ diagram. The sources from the 
cluster and control regions are represented by black filled 
circles, and blue filled squares, respectively.
The locus for class {\sc v} stars at 3.2 kpc and the positions corresponding to 
each spectral type from A0{\sc v} till O8{\sc v},
are indicated by the black continuum line. 
The reddening vector representing the standard interstellar reddening law (Fitspatrick, 1999), is 
indicated by the black dotted line.
We also indicate by the gray 
line, the main-sequence reddened by $A_V$=4.5 magnitudes. 
The diagram also presents the 2 and 3 Myear PMS tracks (represented by the 
red and orange continuous lines, respectively), computed for masses in the range 0.1-7
M$_\odot$ (Siess et al. 2000), distance 
of 3.2 kpc and interstellar absorption corresponding to $A_V$=4.5 mag.}

         \label{FigVibStab} 
   \end{figure*}
   
Useful information about the nature of the sources detected in the cluster region, can be obtained from 
the $J-H$ versus $H-K_s$ 
colour-colour (CC) diagram, shown in Fig.6.
There we represented the non-reddened locus for main sequence, giants 
(Koornneef, 1983), and classical T-Tauri (CTT) stars (Meyer et al., 1997), together with the associated 
reddening vector lines, 
which follow the standard interstellar law taken from Fitzpatrick (1999).
We can see that the bulk of the cluster sources are concentrated in
a small region of the CC diagram, presenting colours in the range 0.95 $\leq$ $J-H$ 
$\leq$ 1.4 and 0.35 $\leq$ $H-K$ $\leq$ 0.9. The majority of the objects in this group 
are concentrated along the reddening vector band 
of the empirical CTT locus, indicating that the cluster stellar population is formed 
by a significant number of sources still in their pre-main sequence (PMS) evolutionary phase.
This sources are identified in the last column of Table I as "CTT candidate". 

\subsubsection{Contamination by field stars}

The contamination by Galactic field sources in the cluster stellar population,
can be evaluate taking into account the magnitudes and colours of the sources 
detected in the control region. 

From Fig.6, we notice that there is a group of sources occupying the lower-left part of the CC diagram, 
presenting colours in the range 0.2 $\leq$ $J-H$ $\leq$ 0.6 and 0.1 $\leq$ $H-K$ $\leq$ 0.3.
Some of these objects can be true members of the cluster
(the main ionizing source in the region belongs to this group), however in this particular case, 
the contamination by field stellar sources may be 
important. In fact, their location in the CC space coincides with that occupied by the majority of the sources 
detected in the control field (represented by the blue filled squares), which 
have small colours, as expected for a foreground Galactic stellar population. On the other hand,
we can see from Fig.7 that the majority of the objects in the control region 
present colours in the range 0.1 $\leq$ $(H-K_S)$ $\leq$ 0.35, and magnitudes 
$J$ $\geq$ 14.5. On the other hand, we also notice that many objects present $H-K_S$ colours excess less than
that for the reddened main sequence locus, indicating that they probably 
do not belong to the star formation region.
The majority of the sources in this group 
are referred in Table I as "field source", indicating their status of probable non-cluster members.

\subsubsection{Cluster mean visual extinction}

Assuming that the sources presenting colours in the range 0.95 $\leq$ $J-H$ 
$\leq$ 1.4 and 0.35 $\leq$ $H-K$ $\leq$ 0.7 are CTT stars, 
and remembering that the T-Tauri locus was derived from modeling accretion
disks (Meyer et al. 1997), we can estimate a value for
the \emph{mean visual extinction} to this group of cluster sources.
From the interstellar extinction law of Fitspatrick (1999), we applied the corresponding 
de-reddening vector to each source in these group, and 
found a mean value for $A_V$ of about 4.5$\pm$1.5 magnitudes. We can evaluate this 
result representing in Fig.6 (by red lines) the CTT locus reddened by 3, 4.5 and 
6 magnitudes of visual extinction. One can see that the majority of the sources lie 
well in between the two extreme reddened CTT locus, which indicates
that the extinction to the cluster members ranges between $A_V$=3, and $A_V$=6 magnitudes.
The mean value $A_V$= 4.5 mag is somewhat higher than that found from the 
photometry of the IRS1 source. The reason for this small difference probably is due to the fact that
the dust destruction process (generated by UV photons of the massive star),
is more effective in the innermost part of the cluster, resulting in a lower
visual extinction. This assumption is corroborate by the composite three colour image of the cluster core, shown in Fig.4.
In fact, we can see that the central part of the cluster appears "cleaned" while other 
sources in the region seems to be involved by a tenuous extended emission, which probably is produced 
by scattered stellar light. 
Indeed, both results agree well when we take into account the associated uncertainty.

\begin{figure*}
\centering
  \vspace{5pt}
   \includegraphics[width=14cm,clip]{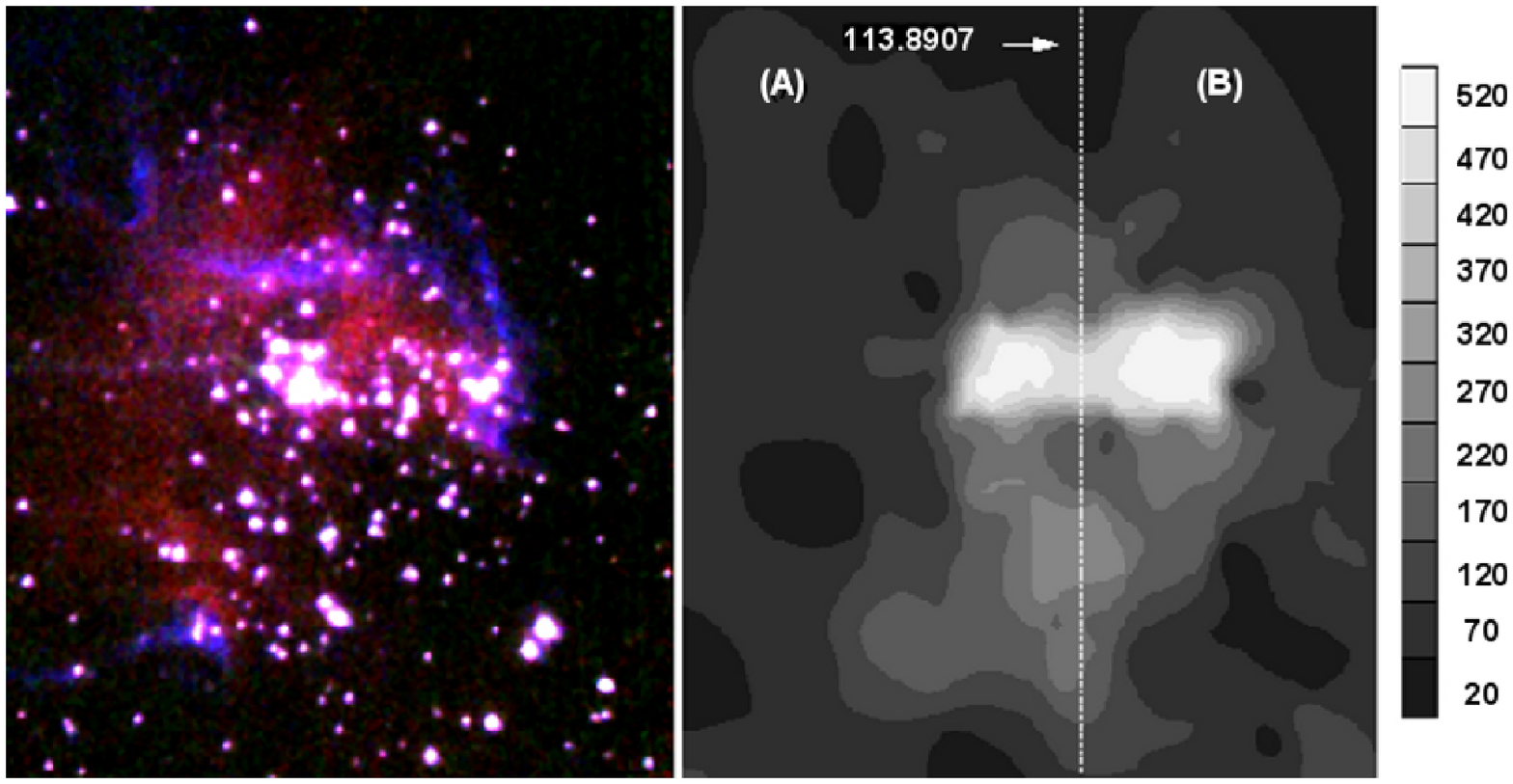}
      \caption{(left) The three colour composite image made from the Br$\gamma$ (red),
	  Continuum (Green) and H$_2$ (Blue) frames. North is to the Top, east to the
	  left, and the field of view is about 70\arcsec$\times$60\arcsec.
	  (right) The magnitude limited ($K_S<$15.5) surface density distribution of sources in the cluster region. The contours start from a surface density of 20 sources pc$^{-2}$ till a maximun of 520 sources pc$^{-2}$. The white dotted line divide the cluster field in two regions (A) and (B), in which is performed the study of excess emission sources.
	  }
         \label{FigVibStab}
   \end{figure*}

   \subsubsection{The colour-magnitude diagram}
 
Additional information about the 
stellar population in the direction of the Sh 2-307 region can be obtained using the \textbf{$K_S$ versus $H-K_S$} 
colour-magnitude (CM) diagram. In Fig.7, we show the diagram 
constructed from the photometric data of the sources in Table 1.
In all diagrams, the sources in the cluster region (that 
found simultaneously in the $J$, $H$ and $K_S$ bands), are represented by the black filled circles. 
As in the case of the CC diagram, the contamination by field stars in the cluster population can be evaluate from the 
blue filled squares (representing
the objects detected in the control region). The locus for class {\sc v} stars at 3.2 kpc was plotted, and 
the spectral types from A0{\sc v} till O8{\sc v} are indicated. The
intrinsic colours were taken from Koornneef (1983), while
the absolute $J$ magnitudes were calculated from the absolute
visual luminosity for the zero-age main sequence (ZAMS)
taken from Hanson et al. (1997). The reddening vector for an
O9.5{\sc v} star is
shown by the dashed line, and the positions corresponding to $A_V$=5, 10, 15
and 20 magnitudes (Fitspatrick, 1999), are indicated by crosses. For completeness,
we also indicate by the gray line, the main-sequence locus reddened
by $A_V$ = 4.5 mag, the mean visual extinction inferred from the reddened CTT locus 
in the CC diagram.

It is interesting to notice that some of the excess emission sources are too bright to be individual low-mass and T-Tauri objects. For example, from the brightness and colours of the IRS4 and IRS5 and IRS30 sources, we suspect that as in the case of the IRS2 one, they are candidate to intermediate mass young stellar objects. Unfortunately, from the present dataset this is all we can say about such sources. In this sense, they are excellent targets for further near- and mid-IR spectrophotometric studies.

\subsubsection{Age estimates for the cluster stellar population}

One can determine the approximate age of a cluster population by comparing their observed magnitudes and colours with 
that from evolutionary stellar models. For very young stellar clusters, one method consists in to study the
post-main sequence evolution of the earliest members (if significant evolution has occurred). In this sense,
we can initially assume the main sequence life time of O9{\sc v}-9.5{\sc v} stars of about 5 Myrs (Meynet et al. 1994; Palla \& Stahler 1999)
as an upper limit for the cluster age .
On the other hand, one can also infer the age of the stellar population
by fitting the low-mass pre-main sequence members using theoretical isochrones.
Fitting PMS isochrones to young stellar clusters is still an uncertain process, mainly for
stellar ages less than 1 Myrs (Baraffe et all. 2002). Indeed its usage is certainly useful to obtain 
estimates for the cluster age, and in the case of non-coeval stellar populations, to get an idea of their
\emph{relative} ages.

In order to perform the isochrones fit to the cluster sources, we show in the \textbf{$K_s$} versus $H-K_s$ CM diagrams  (Fig. 7),
the 2 and 3 Myear PMS 
tracks (represented by the red and orange continuous lines, respectively), computed for Z=0.02 
and masses in the range 0.1-7 M$_\odot$ 
(Siess et al. 2000). The \textbf{$K_s$}-band magnitudes were computed for a distance 
of 3.2 kpc and the colours were reddened by $A_V$=4.5 mag (Fitspatrick, 1999),
the mean visual extinction inferred from the CTTS locus (see Fig.6). 
As already discussed in the analyses made for the CC diagram, the contamination by Galactic field stars must be small. The probable non-cluster members
are the low luminosity objects that present magnitude and colours similar to that indicated by the blue square symbols
(representing the sources detected in the control region). The probable exceptions 
are the brightest cluster candidates that are well placed in the core of the cluster region (as is the case of the IRS1 source).

From the comparison of the early-type sources in the CM diagram (0.2 $\leq H-K_s\leq$ 0.5 and \textbf{$K_s\sim 13.5$)} with the PMS evolutionary tracks, 
we can see that the main-sequence (MS) turn-on of the cluster probably is occurring for the 2-3 Myrs old sources corresponding to ZAMS stars of 3-3.5 M$_\odot$.
This is an important information that indicates the probable age of the O star.
Another useful feature that seems to indicate the presence of a 2-3 Myrs old stellar population, comes from an inspection of the low mass end of the cluster source distribution.
We can notice that the oldest PMS tracks "fit" well the left boundary of the
cluster source sample, including the "knee" at the 0.3-0.5 M$_\odot$ range.
On the other hand, we also can see that there are a large quantity of objects farthest to the right (much more than 1.5 magnitudes) of the \emph{reddened} PMS track locus. Such sources present strong excess emission in the K-band, a feature that amplify the broadening of the distribution of these sources in the colour-magnitude space.

Another possible explanation for the broad distribution of sources in the CM space, can be obtained if we assume that the cluster stellar population is formed by sources in different evolutionary stages. In this sense, the observed scatter could be due to a combination of photometric uncertainty, small differential reddening and an age spread of several hundred thousand years. 
Evidence indicating the presence of an age spread on the stellar population may come from the comparison of the 2 Myr PMS model with the low mass end of the source cluster distribution. In this case, we can see that several low luminosity sources are reasonably 
well fitted by the low-mass end of the 2 Myrs PMS track, which however is not true when we consider the brightest and reddest objects. Even the use of a younger PMS model (like the 1 and 0.5 Myrs one) does not result in reasonable fit to the reddest sample and the reason for that probably is the fact that the SEDs predicted by the PMS models result only from the contracting photospheres on the Hayashi tracks (vertical tracks in the CM diagrams), ignoring the effects produced by the presence of acreting disks. 
The NIR excess generated by accretion disks would explain (in principle) why these sources appears red-wards of the predicted PMS isochrones.

An important cluster parameter related with the presence of disks, is the frequency of cluster members that show NIR excess emission, and its relation with the age of the cluster stellar population. In fact, the initial disk frequency and the variation of the cluster-disk frequency with the cluster age, sets the timescale for both, the disk evolution and the lifetime of the circunstellar disk phase (Lada et al. 2000, Lada \& Lada 2003).
Haisch, Lada \& Lada (2001) found that the disk fraction rapidly drops with the cluster age. They conclude that for a one-million year cluster age, the disk fraction is about 80 percent, dropping to only 10 percent for a five million years cluster age. In this sense, the direct measurement of the cluster disk frequency (inferred from the fraction of sources showing excess emission) enable us to estimate the cluster’s age. The fraction obtained from the J-, H- and K-bands observations can be considered a lower limit for the true number of sources with excess emission in a cluster. This is because some fraction of sources that do have accreting disks, may show JHK colours similar to that presented by reddened early- and late-type "normal" stars. On the other hand, virtually all sources in a cluster that have accreting disks, may be identified in a JHKL diagram where this degenerecence is broken (Lada et al. 2000).

We can obtain another estimated for the mean age of the cluster stellar population, by computing the fraction of sources showing excess emission in the \textit{JHK} colour-colour diagram, and comparing the result with those obtained by Lada et al.(2000), and Haisch, Lada \& Lada (2001).
\textbf{In order to be conservative and to avoid any bias generated by incompleteness sample effects (for example due to the lack of completeness correction in some crowded parts of the cluster region), we will consider only the sources located to the right of the redenning vector for early type sources that have $K_S$ magnitudes $<$ 15.5, a value that is more than one magnitude less than the derived completeness limit for the $K_S$ band (16.7).}
There are 86 sources in Table 1 with K$_S$ $<$15.5 (those detected simultaneously in the three NIR bands) with 14 being probable galactic field sources, and 27 presenting K-band excess emission, which corresponds to a fraction of 38$\pm$8\%. If we assume that the majority of the sources in our sample are less massive than a B star (a reasonable assumption considering the large quantity of T-Tauri star candidates found in this cluster), we can use the results obtained by Lada et al.(2000) to estimate (from their Table 2) as about 60\% the probable true fraction of excess emission sources in this cluster, which corresponds to an age of 2.5 Myrs (Haisch, Lada \& Lada 2001), in good agreement with the value derived from the use of PMS evolutionary tracks in the colour-magnitude diagram.

\subsubsection{Evidence for triggered star formation}

Triggered star formation is a process powered by ionization and/or shock fronts 
generated by massive stars. It is 
believed to occurs at the interface between the H{\sc ii} region and associated molecular cloud, and 
in this context, two main mechanisms have been proposed: The \emph{radiation-driven implosion} and the \emph{collect-and-collapse} models. 
In the first model, the expanding ionization front compresses the existing molecular clumps, 
leading to density enhancements, which eventually exceeds the local critical mass,
collapsing to form new stars (Lefloch \& Lazareff 1994). In the collect-and-collapse model, the neutral
material accumulates between the ionization front of the H{\sc ii}
region and the shock front in the neutral gas. This compressed
material may be dynamically unstable on a short internal
crossing timescale, forming cometary globules, which eventually will lead to the formation of 
massive fragments forming new generations of stars and/or clusters (Elmegreen \& Lada 1977; Garcia-Segura \& Franco 1996).

We searched for signs of triggered star formation, by initially studying the observed surface density distribution of sources, and their correlation with the extended emission in the region.
In Figure 8 (left), we present the three colour combined image constructed from the Br$\gamma$ (2.166$\mu$m - red), H$_2$ (2.12$\mu$m - blue) and Continuum (2.14$\mu$m - green) narrow band filter images. Combining these images together in this way allow us to visualise any spatial correlation between the point sources and observed extended emission, as well as, the effects of the presence of the massive star in the surrounding medium, with the H$_2$ emission tracing the photo-dissociation regions (PDRs), and the Br$\gamma$ image tracing the ionized gas.
In the case of the H$_2$ emission, we notice that it is strong to the north and to the west, but is
absent to the east, indicating that in this direction most of the molecular gas probably was already dissociated by the UV photons of the O star. On the other hand, the Br$\gamma$ emission suddenly diminish to the south and southwest, indicating the presence of neutral gas.

The surface density distribution of the cluster sources reveals other interesting features. In Figure 8 (right), we show the magnitude limited ($K_S<15.5$) surface density distribution of objects in the cluster region. We can see two prominent concentrations there. As expected, the O star (far the most massive source in the region) is related to the concentration that lies close to the geometrical centre of the cluster and associated H{\sc ii} region. On the other hand, the striking result is the presence of a second compact group of sources, which shows close spatial correlation with the strong H$_2$ photo-dissociation regions seem to the right of the extended Br$\gamma$ emission, apparently forming a sub-cluster. 
Finally, and as a last comment in this subject, there is a third less prominent sub-structure seem to the south of the main source. It extends till a small group of objects, associated with the H$_2$ emission seem in the left corner of the image in Figure 8 (left).

Interestingly, the features above described, agree with the scenarios predict by the two main triggering star formation process mentioned earlier. In fact, in the regions where the sources appear associated with the H$_2$ extended emission (the second compact group), the star formation would be occurred accordingly to the radiation-driven implosion model, while in the case of the sources that appears more embedded in the parental molecular cloud (like the sources associated with the third less prominent substructure to the south of IRS1, far away from the H{\sc ii} region ionization front), the collect-and-collapse model seems to be the most probable.

However, in order to properly support this assumption it is necessary to find some evidence for an age spread on the cluster stellar population. We did this by studding the fraction of sources in two cluster areas, as described as follow.
From Figure 8 (right) we computed the fraction of sources showing excess emission in two separate regions (A) and (B), as indicated by the vertical dotted line, which corresponds to the sources found to the left and right from the R.A. coordinate 113.8907 (07:35:33.8).
As already noted before, there are 86 sources in the cluster region presenting $K_S$ magnitudes less than 15.5, with 47 (9 field sources) located in region (A) and 39 (5 field sources) in region (B). The total number of excess sources is 27, 11 from region (A) and 16 from region (B), which correspond to excess emission fractions of 29\% and 47\% respectively.  
This result indicates that the sources in region (B) probably are younger (as a sample) than that found in region (A).

The large number of low to intermediate mass YSO and T-Tauri candidates concentrated at specific distances from
the main cluster source, their apparent spatial correlation with the H$_2$ emission, and the evidence of an age spread in the cluster population, are constrains that favor the assumption of triggered star formation in the region.

\subsection{The Sh 2-307 H{\sc ii} region}

From Figure 8(a) one can see that the Br$\gamma$ extended emission has a nearly spherical 
shape centred on IRS1.
We can estimate the size of the Sh 2-307 H{\sc ii} region by measuring the angular diameter of 
the Br$\gamma$ extended emission, which occurs in both, in the north-south and in the east-west directions. It
measures about 60 arcsec that for a distance of 3.2$\pm 0.5$ kpc corresponds to $d$ $\sim$ 0.94$\pm 0.15$ pc. 
 
The numerical density of the ionized gas can be derived from the emission measure, which can be computed from the detected 5 GHz 
flux density and measured angular diameter, assuming that the emitting region has an spherical shape. 
In the calculation, we used the expression for the expected free-free emission from an optically 
thin plasma at wavelength $\lambda$, given by:
\begin{equation}
S_\nu = 5.4\times 10^{-16}g_{ff}(\lambda,T)\Omega E T^{-1/2}e^{-hc/\lambda kT} \; {\rm Jy}
\end{equation}

\noindent
where $E$ is the emission measure (cm$^{-5}$), $\Omega$ is the solid angle of the source and $g_{ff}(\lambda,T)$ is 
the Gaunt factor, which for radio wavelengths can be computed from:

\begin{equation}
g_{ff}(\lambda, T) = \frac{\sqrt{3}}{\pi}\biggl[17.7+\ln
 \biggl(\frac{T^{3/2}\lambda}{c}\biggr)\biggr]
\end{equation}
Assuming $T$=7500K, $D$=3.2 kpc and $S_\nu$=0.73 Jy we found $E= 3.7\times 10^{23}$cm$^{-5}$, which corresponds to a 
mean electron density $n_e$ $\sim$ 550 cm$^{-3}$. It is interesting to notice that the values of the diameter and 
electron density we found are in good agreement with the density-size relation derived by Kim \& Koo (2001).

The diameter we found for the H{\sc ii} region
indicates that Sh 2-307 had already evolved from the ultra-compact 
phase (when d$\leq$0.1 pc), being in a more advanced evolutionary stage.
We can estimate the age of Sh 2-307, assuming that it reached 
its \emph{initial} Strongren radius in a very short time of a few $10^4$ years
(corresponding to a rapidly expansion phase dominated by the R-type shock - Spitzer 1978), and that
after this phase, it has been expanding in a uniform medium owing to 
the pressure difference between the hot ionized gas and the outer cool molecular gas. This corresponds
to a condition where the H{\sc ii} region expansion is governed by a weak D-type ionization front, with a rate of 
expansion that can be estimate using the equation (Spitzer 1978):
\begin{equation}
R_{f}(t) = R_{i}\biggl(1+
 \frac{7C_{II}t}{4R_{i}}\biggr)^{4/7}
\end{equation}
where $R_i$ and $R_f$ are the initial and final values for the Strongren radius, and $C_{II}$ is the sound
speed ($\sim$ 4km/s) in the H{\sc ii} region. In the calculation we assumed that the numerical density 
in the beginning of the ultra-compact phase was about $ 10^5$cm$^{-3}$ (Churchwell, 2002). 
The initial Strongen radius for an O9{\sc v}-9.5{\sc v} star ($N_{Ly}$=7.9$\times 10^{47} s^{-1}$, Martins et al. 2005)
is $R_i\sim$0.012 pc and considering $R_f$=0.47 pc, we found an age of about (1.6$\pm 0.4$)$\times 10^6$ years.
Despite the uncertainty due to the assumption of uniform medium and because the velocity of expansion probably
changes with time, the derived value probably is a reasonable estimation for the age of the H{\sc ii} region, considering that
it had already evolved from the UC phase, and probably is still expanding, as suggested by the presence of PDRs.

From the observed 5GHz flux density we can obtain another independent estimate for the heliocentric distance of the H{\sc ii}
region. In the calculation we used the relation between $N_{Ly}$ and $S_\nu$
given by Urquhart et al. (2004),
\begin{equation}
N_{Ly} = 7.7\times 10^{43}S_{\nu} D^{2} \nu^{0.1}  s^{-1}
\end{equation}
with $S_{\nu}$ in mJy, $D$ in kpc and $\nu$ in GHz.
For $N_{Ly}$=7.9$\times 10^{47} s^{-1}$ and $S_\nu$=730 mJy, we compute a distance of about 3.5 kpc, in good agreement with 
the spectrophotometric result of 3.2$\pm 0.5$ kpc.
This result can be considered an \emph{upper limit} considering that some fraction of the Lyman continuum photons 
produced by the O star certainly are "lost" due to
dust absorption or by escaping from the region.
From this result we estimate as $\sim$ 85 \% the fraction of Lyman continuum photons generated by the
massive star that effectively ionizes the gas in the H{\sc ii} region.

\section{Summary}

In this work we performed a detailed study of the Sh 2-307 H{\sc ii} region, and associated stellar population. 
From the high quality SOAR-OSIRIS NIR broad-band images, we were able to resolve the sources placed in the innermost zones, identifying the ionizing agent of the H{\sc ii} region detected at optical and radio wavelengths.
The analyses of the J-, H- and K-band OSIRIS spectrophotometric data of this object, enable us to classify it as an O9{\sc v}-O9.5{\sc v} star, reddened by $A_V=3.6 \pm 0.7$ magnitudes. It is probably placed at a heliocentric distance of 3.2$\pm$0.5 kpc that for R$_0$ = 8 kpc, puts the Sh 2-307 H{\sc ii} region at more than 10 kpc from the  Galactic centre.

There are 179 NIR objects in an area of about 1.2 $\times$ 1.2 square arcmin, the majority of them 
presenting colours characteristic of T-Tauri sources. 
From the comparison of the early-type sources in the CM diagram with the PMS evolutionary tracks, we
found that the main-sequence turn-on is probably occurring for sources 2-3 Myrs old, which correspond to 3-3.5 M$_\odot$ ZAMS stars.
Considering the brightness and colours of the IRS2, IRS4 and IRS5 sources, we suspect that they probably are intermediate mass young stellar objects.
A constrain that provides some support for this assumption, is the detection of CO band absorption lines in the NIR spectra of the IRS2 source, a feature commonly found in intermediate-mass YSOs.

The narrow-band Br$\gamma$ image shows that the H{\sc ii} region has an spherical like shape with a mean radius 
of about 0.47$\pm$0.15 pc, indicating that it had already evolved from the ultra-compact phase.
From the 5 GHz flux density taken from the PMN catalogue, and the measured radius, we estimate a mean 
electron density of 550 cm$^{-3}$ and an H{\sc ii} region age of about 1.6$\times 10^6$ years.
The PDRs at the interface between the ionized and molecular gas, are well
traced by the H$_2$ emission detected in
the narrow band images. The spatial correlation
between the ionized gas, and the PDRs (clearly seem to the north and to the west of the ionizing source) 
lead us to conclude that there is a high degree of interaction of the UV photons from the O
star with the nearby molecular cloud, suggesting that the H{\sc ii} region probably is still expanding into the surrounding medium.
On the other hand, there is no H$_2$ emission to the east, indicating the absence of molecular gas in this part of the region. Indeed, we detect strong Br$\gamma$ emission there, and considering its interaction with small H$_2$ emitting "clumps" well seem to the north and to the south, we conclude for the presence of
a "champagne flow" occurring at the border of the parental molecular cloud, powered by the ionization front of the 
associated H{\sc ii} region. This conclusion agrees with the assumption made by Russeil et al. (1995), who suggested that a "champagne flow" in the region would produce the observed discrepancy in the corresponding HII region radial velocity, inferred from their H$\alpha$ Fabry-Perot observations.
Finally, based on the analyses of the relative fraction of excess emission sources in the region, and the observed spatial distribution of low-mass YSO and T-Tauri stars relative to the main cluster source, we conclude that the O star probably has been triggering the star formation process in the region.

\section*{Acknowledgements}

      We would like to thanks the comments and suggestions made by the anonymous referee. Her/his comments were 
	  very useful to improve the quality and presentation of the final manuscript. This work was partially supported 
	  by the ALMA-CONICYT Fund, under the project number 31060004,
	  "A New Astronomer for the Astrophysics Group, Universidad de La Serena", and by the Physics department of the 
	  Universidad de La Serena.  
	  We acknowledge the staff of the SOAR Telescope for their efficient support during the
	  observations.
      This publication makes use of data products from the Two Micron All 
      Sky Survey, which is a joint project of the University of 
      Massachusetts and the Infrared Processing and Analysis Center/California 
      Institute of Technology, funded by the
      National Aeronautics and Space Administration and the National Science Foundation.
	  This work was partially supported by the Brazilian agencie FAPESP.

\begin{table*}
\caption{The list of known OB stars used as templates.}
\label{catalog}
\centering
\renewcommand{\footnoterule}{}  
\begin{tabular}{cc}
\hline \hline
ID & SpType \\

\hline
   HD149438  & B0.2{\sc v}\\
   HD123056  & O9.5{\sc v}\\
   HD93028  & O9{\sc v}\\
   HD165246  & O8{\sc v}\\
   HD152623  & O7{\sc v}\\
\hline
\end{tabular}
\end{table*}

\begin{table*}
\centering
\caption{The $J, H$ and $K_S$ OSIRIS photometry of the sources detected in the direction of SH 2-307 HII region (part 1).}
\label{catalog}
\centering
\renewcommand{\footnoterule}{}  
\begin{tabular}{lccccccc}
\hline \hline
IRS & RA     &  Dec    & J & H & K & Note & NIR K-band excess?\\
~ &(J2000) & (J2000) &~    & \\
\hline

1	&	113,8924	&	-18,7597	&	10,66	$\pm$	0,05	&	10,35	$\pm$	0,06	&	10,21	$\pm$	0,07	&	O9{\sc v}-O9.5{\sc v}	&		\\
2	&	113,8847	&	-18,7669	&	12,60	$\pm$	0,05	&	11,83	$\pm$	0,07	&	11,49	$\pm$	0,09	&	YSO candidate	&		\\
3	&	113,8863	&	-18,7597	&	13,49	$\pm$	0,05	&	12,78	$\pm$	0,06	&	12,49	$\pm$	0,08	&	B2{\sc v}	&		\\
4	&	113,8852	&	-18,7675	&	14,48	$\pm$	0,05	&	13,31	$\pm$	0,06	&	12,58	$\pm$	0,08	&	YSO candidate	&	yes	\\
5	&	113,8913	&	-18,7665	&	15,51	$\pm$	0,05	&	14,01	$\pm$	0,06	&	12,90	$\pm$	0,08	&	YSO candidate	&	yes	\\
6	&	113,8934	&	-18,7593	&	13,75	$\pm$	0,05	&	13,21	$\pm$	0,06	&	12,98	$\pm$	0,08	&	B5{\sc v}	&		\\
7	&	113,8943	&	-18,7629	&	14,39	$\pm$	0,05	&	13,45	$\pm$	0,06	&	13,09	$\pm$	0,08	&	T-Tauri candidate	&		\\
8	&	113,8865	&	-18,7696	&	13,74	$\pm$	0,05	&	13,31	$\pm$	0,06	&	13,13	$\pm$	0,08	&	B5{\sc v}-B8{\sc v}	&		\\
9	&	113,9019	&	-18,7542	&	13,78	$\pm$	0,05	&	13,33	$\pm$	0,06	&	13,20	$\pm$	0,08	&	field source	&		\\
10	&	113,8827	&	-18,7686	&	14,88	$\pm$	0,05	&	13,83	$\pm$	0,06	&	13,35	$\pm$	0,08	&	T-Tauri candidate	&		\\
11	&	113,8918	&	-18,7660	&	14,51	$\pm$	0,05	&	13,71	$\pm$	0,06	&	13,36	$\pm$	0,08	&	field source	&		\\
12	&	113,8932	&	-18,7585	&	15,01	$\pm$	0,05	&	13,93	$\pm$	0,06	&	13,40	$\pm$	0,08	&	T-Tauri candidate	&		\\
13	&	113,8885	&	-18,7706	&	14,31	$\pm$	0,05	&	13,83	$\pm$	0,06	&	13,56	$\pm$	0,08	&	T-Tauri candidate	&		\\
14	&	113,8965	&	-18,7645	&	15,24	$\pm$	0,05	&	14,10	$\pm$	0,06	&	13,68	$\pm$	0,08	&	T-Tauri candidate	&		\\
15	&	113,8933	&	-18,7637	&	16,06	$\pm$	0,05	&	14,76	$\pm$	0,06	&	13,90	$\pm$	0,08	&	T-Tauri candidate	&	yes	\\
16	&	113,8918	&	-18,7641	&	15,67	$\pm$	0,05	&	14,52	$\pm$	0,06	&	13,91	$\pm$	0,08	&	T-Tauri candidate	&		\\
17	&	113,8915	&	-18,7599	&	15,33	$\pm$	0,08	&	14,42	$\pm$	0,07	&	13,93	$\pm$	0,08	&	T-Tauri candidate	&		\\
18	&	113,8874	&	-18,7589	&	15,72	$\pm$	0,05	&	14,52	$\pm$	0,06	&	13,94	$\pm$	0,08	&	T-Tauri candidate	&		\\
19	&	113,8951	&	-18,7598	&	14,95	$\pm$	0,05	&	14,38	$\pm$	0,06	&	13,96	$\pm$	0,08	&	T-Tauri candidate	&	yes	\\
20	&	113,8923	&	-18,7574	&	16,02	$\pm$	0,05	&	14,76	$\pm$	0,06	&	13,98	$\pm$	0,08	&	T-Tauri candidate	&	yes	\\
21	&	113,8889	&	-18,7604	&	15,70	$\pm$	0,05	&	14,44	$\pm$	0,06	&	14,01	$\pm$	0,08	&	T-Tauri candidate	&		\\
22	&	113,8878	&	-18,7631	&	15,95	$\pm$	0,05	&	14,74	$\pm$	0,06	&	14,03	$\pm$	0,08	&	T-Tauri candidate	&	yes	\\
23	&	113,8922	&	-18,7587	&	15,56	$\pm$	0,06	&	14,53	$\pm$	0,07	&	14,05	$\pm$	0,08	&	T-Tauri candidate	&		\\
24	&	113,8870	&	-18,7596	&	15,58	$\pm$	0,05	&	14,54	$\pm$	0,06	&	14,09	$\pm$	0,08	&	T-Tauri candidate	&		\\
25	&	113,8859	&	-18,7591	&	16,72	$\pm$	0,05	&	15,07	$\pm$	0,06	&	14,18	$\pm$	0,08	&	T-Tauri candidate	&	yes	\\
26	&	113,8881	&	-18,7521	&	14,91	$\pm$	0,05	&	14,42	$\pm$	0,06	&	14,19	$\pm$	0,08	&	field source	&		\\
27	&	113,8929	&	-18,7585	&	15,76	$\pm$	0,05	&	14,74	$\pm$	0,06	&	14,22	$\pm$	0,08	&	T-Tauri candidate	&		\\
28	&	113,8910	&	-18,7609	&	16,24	$\pm$	0,05	&	15,02	$\pm$	0,06	&	14,23	$\pm$	0,08	&	T-Tauri candidate	&	yes	\\
29	&	113,8892	&	-18,7610	&	15,45	$\pm$	0,05	&	14,71	$\pm$	0,06	&	14,25	$\pm$	0,08	&	T-Tauri candidate	&	yes	\\
30	&	113,8937	&	-18,7713	&	16,94	$\pm$	0,05	&	15,44	$\pm$	0,06	&	14,29	$\pm$	0,08	&	YSO candidate	&	yes	\\
31	&	113,8941	&	-18,7637	&	16,00	$\pm$	0,05	&	14,9	$\pm$	0,06	&	14,3	$\pm$	0,08	&	T-Tauri candidate	&		\\
32	&	113,8889	&	-18,7601	&	16,48	$\pm$	0,05	&	15,21	$\pm$	0,06	&	14,33	$\pm$	0,08	&	T-Tauri candidate	&	yes	\\
33	&	113,8924	&	-18,7610	&	16,11	$\pm$	0,05	&	15,02	$\pm$	0,06	&	14,44	$\pm$	0,08	&	T-Tauri candidate	&		\\
34	&	113,8898	&	-18,7600	&	15,44	$\pm$	0,05	&	14,65	$\pm$	0,06	&	14,50	$\pm$	0,08	&	field source	&		\\
35	&	113,8918	&	-18,7586	&	16,65	$\pm$	0,05	&	15,40	$\pm$	0,06	&	14,58	$\pm$	0,08	&	T-Tauri candidate	&	yes	\\
36	&	113,8932	&	-18,7668	&	16,19	$\pm$	0,05	&	15,06	$\pm$	0,06	&	14,62	$\pm$	0,08	&	T-Tauri candidate	&		\\
37	&	113,8939	&	-18,7490	&	16,67	$\pm$	0,05	&	15,27	$\pm$	0,06	&	14,63	$\pm$	0,08	&	field source	&		\\
38	&	113,8948	&	-18,7540	&	16,90	$\pm$	0,05	&	15,53	$\pm$	0,06	&	14,65	$\pm$	0,08	&	T-Tauri candidate	&	yes	\\
39	&	113,8894	&	-18,7550	&	16,18	$\pm$	0,05	&	15,12	$\pm$	0,06	&	14,66	$\pm$	0,08	&	T-Tauri candidate	&		\\
40	&	113,8959	&	-18,7617	&	15,35	$\pm$	0,05	&	14,92	$\pm$	0,06	&	14,70	$\pm$	0,08	&	field source	&		\\
41	&	113,8900	&	-18,7635	&	16,76	$\pm$	0,05	&	15,53	$\pm$	0,06	&	14,71	$\pm$	0,08	&	T-Tauri candidate	&	yes	\\
42	&	113,8811	&	-18,7605	&	16,34	$\pm$	0,05	&	15,24	$\pm$	0,06	&	14,76	$\pm$	0,08	&	T-Tauri candidate	&		\\
43	&	113,8943	&	-18,7606	&	16,11	$\pm$	0,05	&	15,13	$\pm$	0,06	&	14,76	$\pm$	0,08	&	T-Tauri candidate	&		\\
44	&	113,8830	&	-18,7626	&	16,68	$\pm$	0,05	&	15,52	$\pm$	0,06	&	14,81	$\pm$	0,08	&	T-Tauri candidate	&	yes	\\
45	&	113,8884	&	-18,7594	&	16,57	$\pm$	0,05	&	15,46	$\pm$	0,06	&	14,84	$\pm$	0,08	&	T-Tauri candidate	&	yes	\\
46	&	113,8902	&	-18,7639	&	16,32	$\pm$	0,05	&	15,23	$\pm$	0,06	&	14,85	$\pm$	0,08	&	T-Tauri candidate	&		\\
47	&	113,8970	&	-18,7645	&	16,50	$\pm$	0,05	&	15,29	$\pm$	0,06	&	14,88	$\pm$	0,08	&	T-Tauri candidate	&		\\
48	&	113,9015	&	-18,7653	&	15,77	$\pm$	0,05	&	15,06	$\pm$	0,06	&	14,90	$\pm$	0,08	&	field source	&		\\
49	&	113,8924	&	-18,7625	&	17,48	$\pm$	0,05	&	15,98	$\pm$	0,06	&	14,92	$\pm$	0,08	&	T-Tauri candidate	&	yes	\\
50	&	113,8831	&	-18,7564	&	16,79	$\pm$	0,05	&	15,67	$\pm$	0,06	&	14,98	$\pm$	0,08	&	T-Tauri candidate	&	yes	\\
51	&	113,9016	&	-18,7631	&	15,69	$\pm$	0,05	&	15,14	$\pm$	0,06	&	14,99	$\pm$	0,08	&	field source	&		\\
52	&	113,8949	&	-18,7646	&	16,51	$\pm$	0,05	&	15,46	$\pm$	0,06	&	15,01	$\pm$	0,08	&	T-Tauri candidate	&		\\
53	&	113,8928	&	-18,7600	&	&	&	15,03	$\pm$	0,17	&		&		\\
54	&	113,8939	&	-18,7651	&	16,95	$\pm$	0,05	&	15,70	$\pm$	0,06	&	15,03	$\pm$	0,08	&	T-Tauri candidate	&		\\
55	&	113,8929	&	-18,7616	&	16,45	$\pm$	0,05	&	15,46	$\pm$	0,06	&	15,04	$\pm$	0,08	&	T-Tauri candidate	&		\\
56	&	113,8905	&	-18,7541	&	17,09	$\pm$	0,05	&	15,84	$\pm$	0,06	&	15,06	$\pm$	0,08	&	T-Tauri candidate	&	yes	\\
57	&	113,8892	&	-18,7621	&	17,00	$\pm$	0,05	&	15,75	$\pm$	0,06	&	15,06	$\pm$	0,08	&	T-Tauri candidate	&	yes	\\
58	&	113,8890	&	-18,7551	&	16,52	$\pm$	0,05	&	15,48	$\pm$	0,06	&	15,06	$\pm$	0,08	&	T-Tauri candidate	&		\\
59	&	113,8870	&	-18,7611	&	16,46	$\pm$	0,05	&	15,47	$\pm$	0,06	&	15,08	$\pm$	0,08	&	T-Tauri candidate	&		\\
60	&	113,8998	&	-18,7679	&	16,80	$\pm$	0,05	&	15,53	$\pm$	0,06	&	15,09	$\pm$	0,08	&	T-Tauri candidate	&		\\

\hline
\end{tabular}
\end{table*} 

\begin{table*}
\centering
\caption{The $J, H$ and $K_S$ OSIRIS photometry of the sources detected in the direction of SH 2-307 HII region (part 2).}
\label{catalog}
\centering
\renewcommand{\footnoterule}{}  
\begin{tabular}{lccccccc}
\hline \hline
IRS & RA     &  Dec    & J & H & K & Note & NIR K-band excess?\\
~ &(J2000) & (J2000) &~    & \\
\hline

61	&	113,8860	&	-18,7604	&	16,61	$\pm$	0,05	&	15,56	$\pm$	0,06	&	15,10	$\pm$	0,08	&	T-Tauri candidate	&		\\
62	&	113,8881	&	-18,7569	&	16,97	$\pm$	0,05	&	15,82	$\pm$	0,06	&	15,14	$\pm$	0,08	&	T-Tauri candidate	&	yes	\\
63	&	113,8898	&	-18,7595	&	16,96	$\pm$	0,05	&	15,75	$\pm$	0,06	&	15,16	$\pm$	0,08	&	T-Tauri candidate	&		\\
64	&	113,8916	&	-18,7664	&	16,83	$\pm$	0,05	&	15,60	$\pm$	0,06	&	15,19	$\pm$	0,08	&	T-Tauri candidate	&		\\
65	&	113,8821	&	-18,7599	&	17,29	$\pm$	0,05	&	15,98	$\pm$	0,06	&	15,20	$\pm$	0,08	&	T-Tauri candidate	&	yes	\\
66	&	113,8944	&	-18,7596	&	17,05	$\pm$	0,05	&	15,78	$\pm$	0,06	&	15,25	$\pm$	0,08	&	T-Tauri candidate	&		\\
67	&	113,8852	&	-18,7631	&	17,08	$\pm$	0,05	&	15,88	$\pm$	0,06	&	15,27	$\pm$	0,08	&	T-Tauri candidate	&		\\
68	&	113,8865	&	-18,7600	&	16,47	$\pm$	0,06	&	15,51	$\pm$	0,07	&	15,33	$\pm$	0,08	&	field source	&		\\
69	&	113,8959	&	-18,7669	&	&	&	15,33	$\pm$	0,08	&		&		\\
70	&	113,8937	&	-18,7604	&	16,78	$\pm$	0,05	&	15,75	$\pm$	0,06	&	15,33	$\pm$	0,08	&	T-Tauri candidate	&		\\
71	&	113,8932	&	-18,7675	&	16,63	$\pm$	0,05	&	15,61	$\pm$	0,06	&	15,34	$\pm$	0,08	&	field source	&		\\
72	&	113,8873	&	-18,7696	&	17,25	$\pm$	0,05	&	15,99	$\pm$	0,06	&	15,34	$\pm$	0,08	&	T-Tauri candidate	&		\\
73	&	113,8878	&	-18,7654	&	16,90	$\pm$	0,05	&	15,82	$\pm$	0,06	&	15,35	$\pm$	0,08	&	T-Tauri candidate	&		\\
74	&	113,8979	&	-18,7693	&	17,83	$\pm$	0,05	&	16,32	$\pm$	0,06	&	15,38	$\pm$	0,08	&	T-Tauri candidate	&	yes	\\
75	&	113,8915	&	-18,7640	&	16,67	$\pm$	0,05	&	15,63	$\pm$	0,06	&	15,39	$\pm$	0,08	&	field source	&		\\
76	&	113,8935	&	-18,7674	&	17,21	$\pm$	0,05	&	15,96	$\pm$	0,06	&	15,39	$\pm$	0,08	&	T-Tauri candidate	&		\\
77	&	113,8912	&	-18,7600	&	16,78	$\pm$	0,06	&	15,92	$\pm$	0,06	&	15,40	$\pm$	0,08	&	T-Tauri candidate	&	yes	\\
78	&	113,8937	&	-18,7591	&	17,69	$\pm$	0,08	&	16,11	$\pm$	0,07	&	15,41	$\pm$	0,08	&	T-Tauri candidate	&		\\
79	&	113,8855	&	-18,7528	&	17,38	$\pm$	0,05	&	16,20	$\pm$	0,06	&	15,41	$\pm$	0,08	&	T-Tauri candidate	&	yes	\\
80	&	113,8847	&	-18,7625	&	17,65	$\pm$	0,05	&	16,35	$\pm$	0,06	&	15,43	$\pm$	0,08	&	T-Tauri candidate	&	yes	\\
81	&	113,8958	&	-18,7522	&	16,99	$\pm$	0,05	&	15,94	$\pm$	0,06	&	15,43	$\pm$	0,08	&	T-Tauri candidate	&		\\
82	&	113,8876	&	-18,7627	&	17,65	$\pm$	0,05	&	16,28	$\pm$	0,06	&	15,45	$\pm$	0,08	&	T-Tauri candidate	&	yes	\\
83	&	113,8890	&	-18,7652	&	16,51	$\pm$	0,05	&	15,72	$\pm$	0,06	&	15,46	$\pm$	0,08	&	field source	&		\\
84	&	113,8925	&	-18,7674	&	17,35	$\pm$	0,05	&	16,13	$\pm$	0,06	&	15,46	$\pm$	0,08	&	T-Tauri candidate	&		\\
85	&	113,8959	&	-18,7666	&	17,73	$\pm$	0,06	&	16,19	$\pm$	0,07	&	15,47	$\pm$	0,08	&	T-Tauri candidate	&		\\
86	&	113,8924	&	-18,7585	&	16,78	$\pm$	0,06	&	15,75	$\pm$	0,07	&	15,47	$\pm$	0,08	&	field source	&		\\
87	&	113,8954	&	-18,7668	&	17,05	$\pm$	0,05	&	15,93	$\pm$	0,06	&	15,48	$\pm$	0,08	&	T-Tauri candidate	&		\\
88	&	113,8807	&	-18,7653	&	16,58	$\pm$	0,05	&	15,75	$\pm$	0,06	&	15,48	$\pm$	0,08	&	field source	&		\\
89	&	113,8907	&	-18,7561	&	17,02	$\pm$	0,05	&	16,02	$\pm$	0,06	&	15,52	$\pm$	0,08	&	T-Tauri candidate	&		\\
90	&	113,8883	&	-18,7588	&	17,33	$\pm$	0,05	&	16,18	$\pm$	0,06	&	15,53	$\pm$	0,08	&	T-Tauri candidate	&	yes	\\
91	&	113,8911	&	-18,7688	&	17,15	$\pm$	0,05	&	16,02	$\pm$	0,06	&	15,56	$\pm$	0,08	&	T-Tauri candidate	&		\\
92	&	113,8827	&	-18,7694	&	17,14	$\pm$	0,05	&	16,01	$\pm$	0,06	&	15,58	$\pm$	0,08	&	T-Tauri candidate	&		\\
93	&	113,8905	&	-18,7601	&	17,64	$\pm$	0,07	&	16,56	$\pm$	0,07	&	15,59	$\pm$	0,08	&	T-Tauri candidate	&	yes	\\
94	&	113,8953	&	-18,7572	&	16,97	$\pm$	0,05	&	15,95	$\pm$	0,06	&	15,60	$\pm$	0,08	&	field source	&		\\
95	&	113,8939	&	-18,7534	&	17,73	$\pm$	0,05	&	16,32	$\pm$	0,06	&	15,61	$\pm$	0,08	&	T-Tauri candidate	&		\\
96	&	113,8936	&	-18,7566	&	17,07	$\pm$	0,05	&	16,07	$\pm$	0,06	&	15,63	$\pm$	0,08	&	T-Tauri candidate	&		\\
97	&	113,8829	&	-18,7548	&	17,32	$\pm$	0,05	&	16,22	$\pm$	0,06	&	15,65	$\pm$	0,08	&	T-Tauri candidate	&		\\
98	&	113,8890	&	-18,7587	&	17,20	$\pm$	0,06	&	16,05	$\pm$	0,06	&	15,65	$\pm$	0,08	&	T-Tauri candidate	&		\\
99	&	113,9009	&	-18,7542	&	16,82	$\pm$	0,05	&	15,93	$\pm$	0,06	&	15,65	$\pm$	0,08	&	field source	&		\\
100	&	113,8872	&	-18,7634	&	17,51	$\pm$	0,05	&	16,29	$\pm$	0,06	&	15,66	$\pm$	0,08	&	T-Tauri candidate	&		\\
101	&	113,8916	&	-18,7554	&	18,07	$\pm$	0,06	&	16,63	$\pm$	0,06	&	15,68	$\pm$	0,08	&	T-Tauri candidate	&	yes	\\
102	&	113,8876	&	-18,7609	&	17,64	$\pm$	0,05	&	16,43	$\pm$	0,06	&	15,69	$\pm$	0,08	&	T-Tauri candidate	&	yes	\\
103	&	113,8946	&	-18,7667	&	17,28	$\pm$	0,05	&	16,23	$\pm$	0,06	&	15,69	$\pm$	0,08	&	T-Tauri candidate	&		\\
104	&	113,9020	&	-18,7509	&	17,51	$\pm$	0,05	&	16,25	$\pm$	0,06	&	15,70	$\pm$	0,08	&	T-Tauri candidate	&		\\
105	&	113,8910	&	-18,7522	&	17,22	$\pm$	0,05	&	16,13	$\pm$	0,06	&	15,71	$\pm$	0,08	&	T-Tauri candidate	&		\\
106	&	113,8873	&	-18,7648	&	17,68	$\pm$	0,05	&	16,43	$\pm$	0,06	&	15,73	$\pm$	0,08	&	T-Tauri candidate	&	yes	\\
107	&	113,8936	&	-18,7581	&	17,61	$\pm$	0,09	&	16,23	$\pm$	0,07	&	15,74	$\pm$	0,08	&	T-Tauri candidate	&		\\
108	&	113,8923	&	-18,7558	&	17,45	$\pm$	0,05	&	16,32	$\pm$	0,06	&	15,76	$\pm$	0,08	&	T-Tauri candidate	&		\\
109	&	113,8916	&	-18,7644	&	17,53	$\pm$	0,05	&	16,34	$\pm$	0,06	&	15,77	$\pm$	0,08	&	T-Tauri candidate	&		\\
110	&	113,8885	&	-18,7677	&	17,35	$\pm$	0,05	&	16,26	$\pm$	0,06	&	15,78	$\pm$	0,08	&	T-Tauri candidate	&		\\
111	&	113,8914	&	-18,7698	&	17,39	$\pm$	0,05	&	16,30	$\pm$	0,06	&	15,81	$\pm$	0,08	&	T-Tauri candidate	&		\\
112	&	113,8951	&	-18,7702	&	17,42	$\pm$	0,05	&	16,44	$\pm$	0,06	&	15,82	$\pm$	0,08	&	T-Tauri candidate	&	yes	\\
113	&	113,8891	&	-18,7606	&	17,60	$\pm$	0,06	&	16,23	$\pm$	0,07	&	15,82	$\pm$	0,08	&	T-Tauri candidate	&		\\
114	&	113,8831	&	-18,7695	&	17,41	$\pm$	0,05	&	16,34	$\pm$	0,06	&	15,83	$\pm$	0,08	&	T-Tauri candidate	&		\\
115	&	113,8952	&	-18,7632	&	17,85	$\pm$	0,06	&	16,56	$\pm$	0,06	&	15,83	$\pm$	0,08	&	T-Tauri candidate	&	yes	\\
116	&	113,8893	&	-18,7584	&	16,81	$\pm$	0,05	&	15,96	$\pm$	0,06	&	15,83	$\pm$	0,08	&	field source	&		\\
117	&	113,8915	&	-18,7562	&	17,28	$\pm$	0,05	&	16,25	$\pm$	0,06	&	15,84	$\pm$	0,08	&	T-Tauri candidate	&		\\
118	&	113,8919	&	-18,7601	&		&		&	15,84	$\pm$	0,22	&		&		\\
119	&	113,8895	&	-18,7632	&	17,82	$\pm$	0,05	&	16,60	$\pm$	0,06	&	15,88	$\pm$	0,08	&	T-Tauri candidate	&	yes	\\
120	&	113,8937	&	-18,7609	&	18,31	$\pm$	0,07	&	16,85	$\pm$	0,06	&	15,90	$\pm$	0,08	&	T-Tauri candidate	&	yes	\\

\hline
\end{tabular}
\end{table*} 

\begin{table*}
\centering
\caption{The $J, H$ and $K_S$ OSIRIS photometry of the sources detected in the direction of SH 2-307 HII region (part 3).}
\label{catalog}
\centering
\renewcommand{\footnoterule}{}  
\begin{tabular}{lccccccc}
\hline \hline
IRS & RA     &  Dec    & J & H & K & Note & NIR K-band excess?\\
~ &(J2000) & (J2000) &~    & \\
\hline

121	&	113,8863	&	-18,7669	&	17,50	$\pm$	0,05	&	16,55	$\pm$	0,06	&	15,99	$\pm$	0,08	&	T-Tauri candidate	&	yes	\\
122	&	113,8862	&	-18,7643	&	16,85	$\pm$	0,05	&	16,29	$\pm$	0,06	&	16,02	$\pm$	0,08	&	field source	&		\\
123	&	113,8910	&	-18,7648	&	17,64	$\pm$	0,05	&	16,53	$\pm$	0,06	&	16,02	$\pm$	0,08	&	T-Tauri candidate	&		\\
124	&	113,8919	&	-18,7671	&	18,09	$\pm$	0,06	&	16,75	$\pm$	0,06	&	16,04	$\pm$	0,08	&	T-Tauri candidate	&		\\
125	&	113,8891	&	-18,7544	&	17,64	$\pm$	0,05	&	16,52	$\pm$	0,06	&	16,04	$\pm$	0,08	&	T-Tauri candidate	&		\\
126	&	113,8879	&	-18,7597	&	17,69	$\pm$	0,05	&	16,68	$\pm$	0,07	&	16,1	$\pm$	0,08	&	T-Tauri candidate	&	yes	\\
127	&	113,8883	&	-18,7571	&	17,89	$\pm$	0,06	&	16,66	$\pm$	0,06	&	16,13	$\pm$	0,08	&	T-Tauri candidate	&		\\
128	&	113,8888	&	-18,7538	&	17,81	$\pm$	0,05	&	16,65	$\pm$	0,06	&	16,14	$\pm$	0,08	&	T-Tauri candidate	&		\\
129	&	113,8879	&	-18,7601	&	17,66	$\pm$	0,06	&	16,49	$\pm$	0,07	&	16,14	$\pm$	0,08	&	field source	&		\\
130	&	113,8867	&	-18,7621	&	17,96	$\pm$	0,06	&	16,79	$\pm$	0,06	&	16,14	$\pm$	0,08	&	T-Tauri candidate	&	yes	\\
131	&	113,9014	&	-18,7521	&	17,95	$\pm$	0,05	&	16,73	$\pm$	0,06	&	16,17	$\pm$	0,08	&	T-Tauri candidate	&		\\
132	&	113,8828	&	-18,7643	&	17,73	$\pm$	0,05	&	16,62	$\pm$	0,06	&	16,19	$\pm$	0,08	&	T-Tauri candidate	&		\\
133	&	113,8882	&	-18,7618	&	18,05	$\pm$	0,06	&	16,83	$\pm$	0,07	&	16,2	$\pm$	0,08	&	T-Tauri candidate	&		\\
134	&	113,8868	&	-18,7589	&	18,21	$\pm$	0,07	&	16,92	$\pm$	0,08	&	16,21	$\pm$	0,08	&	T-Tauri candidate	&	yes	\\
135	&	113,8947	&	-18,7649	&	18,14	$\pm$	0,06	&	16,89	$\pm$	0,06	&	16,22	$\pm$	0,08	&	T-Tauri candidate	&	yes	\\
136	&	113,8864	&	-18,7513	&	17,68	$\pm$	0,05	&	16,63	$\pm$	0,06	&	16,23	$\pm$	0,08	&	T-Tauri candidate	&		\\
137	&	113,8939	&	-18,7667	&	17,64	$\pm$	0,05	&	16,67	$\pm$	0,06	&	16,23	$\pm$	0,08	&	T-Tauri candidate	&		\\
138	&	113,8840	&	-18,7703	&	17,75	$\pm$	0,05	&	16,67	$\pm$	0,06	&	16,24	$\pm$	0,08	&	T-Tauri candidate	&		\\
139	&	113,8945	&	-18,7526	&	17,80	$\pm$	0,05	&	16,68	$\pm$	0,06	&	16,26	$\pm$	0,08	&	T-Tauri candidate	&		\\
140	&	113,8945	&	-18,7605	&	17,45	$\pm$	0,06	&	16,53	$\pm$	0,06	&	16,28	$\pm$	0,08	&	field source	&		\\
141	&	113,9000	&	-18,7664	&	17,85	$\pm$	0,05	&	16,70	$\pm$	0,06	&	16,28	$\pm$	0,08	&	T-Tauri candidate	&		\\
142	&	113,8907	&	-18,7627	&	17,97	$\pm$	0,05	&	16,99	$\pm$	0,06	&	16,39	$\pm$	0,08	&	T-Tauri candidate	&	yes	\\
143	&	113,8860	&	-18,7649	&	18,36	$\pm$	0,06	&	17,18	$\pm$	0,07	&	16,40	$\pm$	0,08	&	T-Tauri candidate	&	yes	\\
144	&	113,8817	&	-18,7627	&	17,99	$\pm$	0,05	&	16,87	$\pm$	0,06	&	16,40	$\pm$	0,08	&	T-Tauri candidate	&		\\
145	&	113,8808	&	-18,7579	&	17,68	$\pm$	0,06	&	16,75	$\pm$	0,06	&	16,41	$\pm$	0,08	&	field source	&		\\
146	&	113,8921	&	-18,7629	&	18,41	$\pm$	0,06	&	17,2	$\pm$	0,07	&	16,41	$\pm$	0,08	&	T-Tauri candidate	&	yes	\\
147	&	113,8910	&	-18,7601	&	&	17,18	$\pm$	0,08	&	16,41	$\pm$	0,08	&		&		\\
148	&	113,8900	&	-18,7705	&	17,90	$\pm$	0,05	&	16,88	$\pm$	0,06	&	16,41	$\pm$	0,08	&	T-Tauri candidate	&		\\
149	&	113,8960	&	-18,7589	&	18,41	$\pm$	0,06	&	17,16	$\pm$	0,07	&	16,42	$\pm$	0,08	&	T-Tauri candidate	&	yes	\\
150	&	113,8920	&	-18,7689	&	18,53	$\pm$	0,06	&	17,06	$\pm$	0,06	&	16,43	$\pm$	0,08	&	T-Tauri candidate	&		\\
151	&	113,8804	&	-18,7533	&	17,80	$\pm$	0,06	&	16,93	$\pm$	0,07	&	16,45	$\pm$	0,08	&	T-Tauri candidate	&	yes	\\
152	&	113,8982	&	-18,7578	&	18,27	$\pm$	0,06	&	17,03	$\pm$	0,06	&	16,47	$\pm$	0,08	&	T-Tauri candidate	&		\\
153	&	113,8959	&	-18,7671	&	17,58	$\pm$	0,06	&	16,68	$\pm$	0,06	&	16,49	$\pm$	0,08	&	field source	&		\\
154	&	113,8823	&	-18,7618	&	18,09	$\pm$	0,06	&	17,03	$\pm$	0,06	&	16,51	$\pm$	0,08	&	T-Tauri candidate	&		\\
155	&	113,8889	&	-18,7634	&	18,48	$\pm$	0,06	&	17,20	$\pm$	0,07	&	16,54	$\pm$	0,08	&	T-Tauri candidate	&		\\
156	&	113,8902	&	-18,7614	&	18,25	$\pm$	0,07	&	17,01	$\pm$	0,07	&	16,54	$\pm$	0,08	&	T-Tauri candidate	&		\\
157	&	113,8898	&	-18,7587	&	18,05	$\pm$	0,06	&	17,04	$\pm$	0,07	&	16,56	$\pm$	0,08	&	T-Tauri candidate	&		\\
158	&	113,8976	&	-18,7590	&	18,28	$\pm$	0,06	&	17,04	$\pm$	0,06	&	16,57	$\pm$	0,08	&	T-Tauri candidate	&		\\
159	&	113,8877	&	-18,7527	&	18,03	$\pm$	0,05	&	16,96	$\pm$	0,06	&	16,58	$\pm$	0,08	&	field source	&		\\
160	&	113,8948	&	-18,7658	&	18,67	$\pm$	0,08	&	17,35	$\pm$	0,07	&	16,61	$\pm$	0,08	&	T-Tauri candidate	&	yes	\\
161	&	113,8835	&	-18,7561	&	18,33	$\pm$	0,06	&	17,27	$\pm$	0,07	&	16,62	$\pm$	0,08	&	T-Tauri candidate	&	yes	\\
162	&	113,9018	&	-18,7692	&	18,58	$\pm$	0,07	&	17,36	$\pm$	0,07	&	16,62	$\pm$	0,08	&	T-Tauri candidate	&	yes	\\
163	&	113,8873	&	-18,7524	&	18,10	$\pm$	0,06	&	17,09	$\pm$	0,07	&	16,63	$\pm$	0,08	&	T-Tauri candidate	&		\\
164	&	113,8827	&	-18,7706	&	18,18	$\pm$	0,06	&	17,13	$\pm$	0,06	&	16,63	$\pm$	0,08	&	T-Tauri candidate	&		\\
165	&	113,8867	&	-18,7650	&	18,25	$\pm$	0,06	&	17,07	$\pm$	0,06	&	16,64	$\pm$	0,08	&	T-Tauri candidate	&		\\
166	&	113,8824	&	-18,7649	&	18,11	$\pm$	0,06	&	17,09	$\pm$	0,06	&	16,66	$\pm$	0,08	&	T-Tauri candidate	&		\\
167	&	113,8959	&	-18,7664	&	18,36	$\pm$	0,07	&	17,20	$\pm$	0,07	&	16,68	$\pm$	0,08	&	T-Tauri candidate	&		\\
168	&	113,8983	&	-18,7638	&	18,63	$\pm$	0,07	&	17,43	$\pm$	0,07	&	16,76	$\pm$	0,08	&	T-Tauri candidate	&	yes	\\
169	&	113,8894	&	-18,7702	&	18,86	$\pm$	0,08	&	17,72	$\pm$	0,07	&	16,77	$\pm$	0,08	&	T-Tauri candidate	&	yes	\\
170	&	113,8907	&	-18,7644	&	17,74	$\pm$	0,05	&	17,06	$\pm$	0,06	&	16,78	$\pm$	0,08	&	field source	&		\\
171	&	113,8860	&	-18,7622	&	18,52	$\pm$	0,07	&	17,35	$\pm$	0,07	&	16,82	$\pm$	0,08	&	T-Tauri candidate	&		\\
172	&	113,9023	&	-18,7579	&	18,50	$\pm$	0,06	&	17,39	$\pm$	0,07	&	16,85	$\pm$	0,08	&	T-Tauri candidate	&		\\
173	&	113,8917	&	-18,7686	&	18,82	$\pm$	0,08	&	17,64	$\pm$	0,07	&	16,86	$\pm$	0,08	&	T-Tauri candidate	&	yes	\\
174	&	113,8943	&	-18,7661	&	18,63	$\pm$	0,07	&	17,41	$\pm$	0,07	&	16,88	$\pm$	0,08	&	T-Tauri candidate	&		\\
175	&	113,8977	&	-18,7671	&	18,78	$\pm$	0,07	&	17,60	$\pm$	0,07	&	16,88	$\pm$	0,08	&	T-Tauri candidate	&	yes	\\
176	&	113,8867	&	-18,7549	&	18,96	$\pm$	0,08	&	17,64	$\pm$	0,07	&	16,98	$\pm$	0,08	&	T-Tauri candidate	&		\\
177	&	113,8882	&	-18,7623	&	18,50	$\pm$	0,07	&	17,57	$\pm$	0,08	&	16,98	$\pm$	0,08	&	T-Tauri candidate	&	yes	\\
178	&	113,8976	&	-18,7640	&	18,64	$\pm$	0,07	&	17,46	$\pm$	0,07	&	16,98	$\pm$	0,08	&	T-Tauri candidate	&		\\
179	&	113,8863	&	-18,7570	&	18,72	$\pm$	0,07	&	17,70	$\pm$	0,07	&	17,00	$\pm$	0,08	&	T-Tauri candidate	&	yes	\\

\hline
\end{tabular}
\end{table*}

\end{document}